 \documentclass[11pt]{article}

 \pdfoutput=1   

\usepackage{graphicx}
\usepackage{amsmath}

\usepackage{enumitem}

\usepackage[francais,english]{babel}

\usepackage{float}
\usepackage{ifthen}
\usepackage{color}

\usepackage{hyperref}

\begin{document}


\title{Exergy in meteorology: Definition and properties of moist-air available enthalpy.}

\author{by Pascal Marquet. {\it M\'et\'eo-France.}}


\date{\today}
\maketitle


\vspace*{-10mm}

\begin{center}
{\em Copy of a paper submitted in October 1991 to the \underline{Quarterly Journal of the Royal Meteorological Society}.} \\
{\em Published in Volume 119, Issue 511, pages 567-590, April 1993}: \\
\url{http://onlinelibrary.wiley.com/doi/10.1002/qj.49711951112/abstract}\\
{\em \underline{Corresponding address}: pascal.marquet@meteo.fr}
\end{center}
\vspace{1mm}


\vspace*{-2mm}

\begin{abstract}
The exergy of the dry atmosphere can be considered as another aspect of the meteorological theories of available energies. 
The local and global properties of the dry available enthalpy function, also called flow exergy, were investigated in a previous paper\footnote{\color{blue} Marquet, {\it Q. J. R. Meteorol. Soc.\/}, vol 117, p.449--475 (1991).}

The concept of exergy is well defined in thermodynamics, and several generalizations to chemically reacting systems have already been made. 
Similarly, the concept of moist available enthalpy is presented in this paper in order to generalize the dry available enthalpy to the case of a moist atmosphere. 
It is a local exergy-like function which possesses a simple analytical expression where only two unknown constants are to be determined, a reference temperature and a reference pressure.

The moist available enthalpy, $a_m$, is defined in terms of a moist potential change in total entropy. 
The local function $a_m$ can be separated into temperature, pressure and latent components. 
The latent component is a new component that is not present in the dry case. 
The moist terms have been estimated using a representative cumulus vertical profile. 
It appears that the modifications brought by the moist formulation are important in comparison with the dry case. 
Other local and global properties are also investigated and comparisons are made with some other available energy functions used in thermodynamics and meteorology.
\end{abstract}


 \section{\underline{INTRODUCTION}.} 
\label{section_INTRO}

The basis of the modern concept of exergy was introduced by W. Thomson (1853,
1879), later Lord Kelvin. He defined as motivity the amount of work that a system can
produce by evolving from an initially heterogeneous distribution of temperature to a
constant equilibrium value. Since then the concept of motivity (known as available
energy) has been rediscovered and generalized in many articles and monographs dealing
with thermodynamics (see the historical review in Haywood (1974) or Kestin (1980)).
Nowadays, the available part of the energy is often called the exergy of a system, whereas
its untransformable part is sometimes called the anergy, which corresponds to a dead
state of the same system.

The terms exergy and anergy are not used in atmospheric science, but several
analogous concepts have been introduced in meteorology: 
(i) the available kinetic energy
by Margules (1905); 
(ii) the available potential energies by Lorenz (1955, 1967), Van
Mieghem (1956), Dutton and Johnson (1967), Pearce (1978) and McHall (1990a, 1990b);
(iii) the global static entropic energy by Dutton (1973, 1976) and its local version by
Pichler (1977); and (iv) the entropic potential energy by Blackburn (1983).

A connection between the thermodynamical concept of exergy and some of these
global meteorological approaches (i) to (iv) was proposed in Karlsson (1990) and in
Marquet (1990a, 1991 (collectively referred to as MM hereafter)). 
For instance, it was shown in MM that the local concept of dry available enthalpy
\begin{equation} 
\boxed{ \; a_h \: = \: (h - h_r) \: - \: T_r \: (s - s_r) \;}
\nonumber
\end{equation} 
generalizes some of the results of (i) to (iii). 
This local function $a_h(p, T)$ can be used as
a Lagrangian energy quantity (see list of symbols in the appendix), yielding a Bernoulli
equation verified by $a_h + e_K + e_G$ and leading to new local non-hydrostatic or hydrostatic
energy cycles. 
It is also easy to take into account an uneven topography and dry static
instabilities (i.e. where $\partial s/ \partial z < 0$).

It is explained in MM that, for an isobaric layer of a limited space region, the
expected hydrostatic conversion term 
$- R  \: \overline{\omega} \: \overline{T}/p$ 
and two new boundary fluxes appear.
These terms cannot be obtained by starting with any local approach of the theory of
Lorenz. However, using such a local approach and assuming 
$\overline{\omega} = 0$, Muench (1965),
Brennan and Vincent (1980) and Michaelides (1987) have all mentioned large unbalanced
residuals. Pontaud et al. (1990) have shown that these residuals could correspond to the
new conversion and flux terms obtained with the available enthalpy formulation where
$\overline{\omega} \neq 0$, 
leading to local energy cycles that are more balanced. All these results show that
the introduction of a local exergy-like concept is relevant in meteorology, in agreement
with the local approaches of Pichler (1977) and Karlsson (1990).

Nevertheless, the dry available enthalpy only deals with the energetics of the
atmosphere when considered as a dry ideal gas with parametrized diabatic heating. This
is a crude approximation of the reality, and accordingly Lorenz (1978, 1979) proposed a
moist version of the available energy concept. Livezey and Dutton (1976) also generalized
on the concept of static entropic energy, $T_0 \: \Sigma$, to simple-solution fluid systems (cloudy
air or salt water). As for thermodynamics, Szargut and Styrylska (1969) proposed a
theory of exergetic processes in moist air and Evans (1969, 1980) coined the term 
{\it essergy}
(for {\it essential} aspect of {\it energy}) 
when he studied the exergy of salt water, the principle of
desalination and, more generally, chemically reacting thermodynamic systems. However,
the word essergy has not really spread in thermodynamic literature and so the term
exergy will be used in this paper. In recent years McHall (1991) derived a moist version
of his theory, and Karlsson (1990) investigated the exergy exchanges in the atmosphere
when considered as a fluid mixture with chemical reactions, including numerical investigations
made at the European Centre for Medium-range Weather Forecasts with a
general circulation model.

Similarly, the purpose of this paper is to define an exergy-like local function called
moist available enthalpy and denoted by $a_m$. It is expected to be the natural generalization
of the dry version, $a_h$.

The differences between the various entropic or available energies introduced in
meteorology--see (i) to (iv) above--mainly concern the definition of the reference
state of minimum total potential energy. Using another approach, the essergy function
introduced by Evans (1969) in thermodynamics, and the exergy function studied by
Karlsson (1990) in meteorology, are derived from a fundamental concept of statistical
physics: the ``information gain'', also called ``Kullback function''(Kullback 1959) or
``contrast''. The moist available enthalpy will be introduced differently in this paper,
starting from a moist generalization of the concept of potential change in total entropy
($\Delta S = a_h/T_r$) which is another approach to the dry available enthalpy (see MM).

To begin with, the general thermodynamic assumptions are set out in section 2. The
moist potential change in total entropy is then defined in section 3 together with the
analytical expression of the specific moist available enthalpy function $a_m$ and its various
components. In section 4 a representative vertical profile of a cumulus cloud is used to
estimate the orders of magnitudes of the components of a$a_m$. The thermodynamic equations
and the equations for the moist energy components are reviewed in section 5. A local
moist energy cycle and the local moist Bernoulli equation are then introduced in section
6. Integral properties shown in section 7 lead to possible definitions of $T_r$ and $p_r$ (two
constants to be determined in $a_m$). In section 8 the moist available enthalpy is expressed
in terms of an associated moist potential temperature. Finally, conclusions and remarks
are contained in section 9.

It should be mentioned that most of the following results have appeared first in
French in Marquet (1990b).

 \section{\underline{MOIST THERMODYNAMICS}.} 
\label{section_2}

The general thermodynamic assumptions that are used throughout this study are
based on the thermodynamic local-state theory (De Groot and Mazur 1962; Glansdorff
and Prigogine 1971), together with some common meteorological approximations (see
Hauf and H\"oller (1987), referred to as HH 
hereafter).\footnote{\color{blue} 
The study of papers of De Groot and Mazur (1962), Glansdorff and Prigogine (1971) and Hauf and H\"oller (1987) has been suggested by Jean-Fran\c{c}ois Geleyn and Jean-Fran\c{c}ois Royer in 1990, by very fruitful comments made during oral presentations of the first dry-air results described in Marquet (1990a).}
To enable easy reference these
hypotheses are labeled in this paper by {\boldmath${\cal H}_n$}. 
The notations and symbols used in this paper
have been chosen according to HH.

\begin{itemize}[label=$\bullet$,leftmargin=3mm,parsep=0cm,itemsep=0.1cm,topsep=0cm,rightmargin=2mm]
\vspace*{-1mm}
\item  {\boldmath${\cal H}_1$} {\bf --} 
The atmosphere consists of four components at the same temperature T denoted by
the subscripts or the superscripts $0$ for dry air, $1$ for water vapour, $2$ for liquid water (stable
or supercooled) and $3$ for ice, respectively. 
Precipitation falling at $T'_w$ (the so-called wet-bulb
temperature) cannot explicitly be taken into account.
\item  {\boldmath${\cal H}_2$} {\bf --} 
Ideal gas and pure form assumptions are for ideally mixed gases and condensed
phases respectively. Salt dissolved in drops or ice crystals is ignored.
\item  {\boldmath${\cal H}_3$} {\bf --} 
Local-state theory is valid: the local-state functions defined in ``thermostatic'' (pressure,
temperature, energy, enthalpy, entropy, . . .) remain valid in non-equilibrium
thermodynamics; the first-order Chapman-Enskog development is a good enough
approximation to the distribution function (otherwise entropy would depend explicitly
on gradients of$\rho$ and $T$); variations of $T$ and $\rho$ are ``small enough'' between two
collisions (mean free path); continuum physics can be applied, leading to the concept
of a fluid parcel; gases are not too rarefied (for practical purpose $p > 10^{-1}$~Pa or
$z < 100 $~km); plasmas or shock waves cannot be investigated. For these assumptions
see, for example, Prigogine (1949), Meixner and Reik (1959) and Lebon and Mathieu (1981).
\item  {\boldmath${\cal H}_4$} {\bf --} 
The specific heats at constant volume and pressure ($c_v^j$ and $c_p^j$, $j = 0$ to $3$) together
with the gas constants ($R^i$,  $i= 0$ and $1$) are temperature-independent. 
The normal transition between liquid water and ice occurs at the triple point temperature 
$T_{tr} = 273.16$~K and is pressure-independent. 
Metastable supercooled water can exist.
\item  {\boldmath${\cal H}_5$} {\bf --} 
Under-or super-saturation does not exist anywhere. The saturation pressures denoted
by $p^{i1}( T)$ with $ i = 2$ or $3$ are those over an infinite and plane surface of liquid 
(for $i = 2$) or ice (for $i = 3$).
\item  {\boldmath${\cal H}_6$} {\bf --} 
Thermodynamic and kinematic effects of drop or ice crystal size spectra are not taken
into account.
\item {\boldmath${\cal H}_7$} {\bf --} 
Enthalpy do not depends on pressure for the condensed phases 
($\partial h^2/ \partial p = \partial h^3/ \partial p = 0$). 
The specific volumes of condensed phases are
neglected in the equation of state and in the Clausius-Clapeyron equation: only
specific volumes of gases are considered.
\item  {\boldmath${\cal H}_8$} {\bf --} 
Turbulent fluxes are only molecular ones, although Reynolds terms (the macroscopic
parametrized turbulence) could implicitly be included into the source and sink terms
of the thermodynamic equations.
\item  {\boldmath${\cal H}_9$} {\bf --} 
Kinetic energy of diffusion is not taken into account. This is justified if the time
derivative of the various components with respect to the barycentric motion may be
neglected (see De Groot and Mazur 1962).
\end{itemize}

Any atmospheric parcel is completely determined by the partial pressures
$p^0$ and $p^1$, the temperature $T$ and the concentrations 
$m^k = {\rho}^k / \rho$ ($k = 0$ to $3$) where ${\rho}^k$ and
${\rho} = \sum_k {\rho}^k$ 
are the partial and total densities respectively. 
The mixing ratios of water
components are $r^k = m^k/m^0$ ($k = 1$ to $3$), 
the total mixing ratio of water is $r^t = r^1 + r^2 + r^3$
and the total concentration of water is 
$m^t = m^1 + m^2 + m^3$. 
All these definitions and
notations are chosen according to HH.

In a parcel each component $k = 0$ to $3$ moves with the 
uniform partial velocity ${\bf v}^k$,
and the barycentric velocity v of the parcel is defined by 
${\rho} \; {\bf v} = {\rho}^k \; {\bf v}^k$
(Einstein's summation
convention over Latin or Greek letters is used throughout this paper). 
The material
barycentric time-derivative operator is defined by 
$d/dt = \partial/\partial t + {\bf v} . {\bf \nabla}$, 
where the second term is the barycentric advection.

The change in concentration $dm^k/dt$ can be split into the sum
\begin{equation} 
\frac{d\, m^k}{dt} \; \: = \; \: \frac{d_e \, m^k}{dt}  
                \: + \:  \frac{d_i \, m^k}{dt} 
\: , \label{eq01}
\end{equation} 
where the external ($d_e \, m^k/dt$) and internal 
($d_i \, m^k/dt$) changes in concentration are
\begin{eqnarray} 
\frac{d_e\, m^k}{dt} &=& - \, \frac{1}{\rho}  \: {\bf \nabla} \, . \; {\bf J}^k
\: , \label{eq02} \\
\frac{d_i\, m^k}{dt} &=&  \frac{1}{\rho} \; {\cal J}^k
\: . \label{eq03}
\end{eqnarray} 

The external change $d_e \, m^k/dt$ is created by a transfer 
of mass across the boundaries of the open
parcel.
The internal change 
$d_i \, m^k/dt$ corresponds to the mass produced 
within the parcel.
These changes depend, respectively, on the 
``diffusion flux'' ${\bf J}^k = {\rho}^k \; ({\bf v}^k - {\bf v})$, $k = 0$ to $3$, and the
``convergence of the phase fluxes'' ${\cal J}^k$, $k = 1$ to $3$, 
due to the (chemical) changes of state:
$1 \leftrightarrow 2 \leftrightarrow 3$.

Various local-state functions are either intensive $( p, T )$ or extensive
$(e_i^k, h^k, s^k , ...)$. 
Any extensive function $\psi = m^k \: {\psi}^k$ can be put in 
the alternative form \footnote{\color{blue} 
Notations are the same as in HH.
If the usual subscripts ($d$, $t$, $v$, $l$, $i$) are used to denote dry air, total water, water vapour, liquid water and ice, respectively, (\ref{eq04}) is equivalent to $\psi = q_d \: {\psi}_d \: + \: q_t \: {\psi}_v \: + \: q_l \: ({\psi}_l-{\psi}_v) \: + \: q_i \: ({\psi}_i-{\psi}_v)$, where the specific contents ($q_d$, ..., $q_i$) are the same as the $m^k$ in (\ref{eq04}).}
\begin{equation} 
\psi \; \: = \; \: 
  m^k \: {\psi}^k
     \; \: = \; \: 
  m^0 \: {\psi}^0
        \: + \: 
  m^t \: {\psi}^1
        \: + \: 
  m^2 \: \left( {\psi}^2 - {\psi}^1 \right)
        \: + \: 
  m^3 \: \left( {\psi}^3 - {\psi}^1 \right)
\: , \label{eq04}
\end{equation} 
where $m^t$ is associated with the water vapour component ${\psi}^1$, differently to HH where
$m^t$ is associated with the liquid water component ${\psi}^2$. 
For instance, the specific (namely per unit mass of moist-air) enthalpy 
($h$), entropy ($s$) and internal energy ($e_i$) read
\begin{eqnarray} 
  h  \; \: = \; \:  
   m^k \: h^k
     \; \: = \; \: 
   e_i \: + \: \frac{p}{\rho}
     \; \: = \; \: 
   e_i \: + \: R \: T
     & = &
  m^0 \: h^0
        \: + \: 
  m^t \: h^1
        \: - \: 
  m^2 \: l_{21}
        \: + \: 
  m^3 \: l_{31}
\: , \label{eq05} \\
  s    & = &  
   m^k \: s^k
\: , \label{eq06} \\
  e_i  & = & 
   m^k \: e_i^k
\: , \label{eq07}
\end{eqnarray} 
where $l_{ij} = h^j - h^i$ are the latent heats.\footnote{\color{blue} 
For instance: $L_{\rm vap} = l_{21} = h^1 - h^2 = h_v - h_l$ and $L_{\rm sub} = l_{31} = h^1 - h^3 = h_v - h_i$.}

Using {\boldmath${\cal H}_5$}, the following identity holds whatever the thermodynamic state of the
parcel is, and can help to simplify several formulae (if one of the condensed phases is
present the moist air is saturated and the logarithm is zero, on the other hand if the moist
air is not saturated there is no condensed phase): \footnote{\color{blue} 
In other words: $q_l \: \ln(e/e_{sw}) \: = \: q_i \: \ln(e/e_{si}) \: = \: 0$.}
\begin{equation} 
m^2 \: \ln\left( \frac{p^1}{p^{21}}\right) 
\; \: = \; \: 
m^3 \: \ln\left( \frac{p^1}{p^{31}}\right) 
\; \: = \; \: 
   0
\: . \label{eq08}
\end{equation} 
Using {\boldmath${\cal H}_7$}, the Kirschoff's law ($i, j = 1$, $2$, $3$) and the local Clausius-Clapeyron equation
($i = 2$, $3$) are respectively \footnote{\color{blue} 
In other words: Eq.~(\ref{eq08}) reads $d/dT[\,L_{\rm vap}(T)\,] = c_{pv} - c_l$ and $d/dT[\,L_{\rm sub}(T)\,] = c_{pv} - c_i$ for the Kirschoff's laws;
$d/dT[\,e_{sw}(T)\,] = L_{\rm vap}(T)/[\,R_v \: T^2\,]$  and $d/dT[\,e_{si}(T)\,] = L_{\rm sub}(T)/[\,R_v \: T^2\,]$ for the Clausius-Clapeyron equations.}
\begin{equation} 
\frac{d}{dT}\, \left( l_{ij} \right) 
\; \: = \; \: 
 c^j_p \: -  c^i_p \; ,
\mbox{\hspace*{10mm} and \hspace*{10mm}}
T \; \frac{d}{dT}\, \left( \ln{p^{i1}} \right) 
\; \: = \; \:  \frac{l_{i1}}{R^1 \: T}
\: . \label{eq09}
\end{equation} 
These equations will often be used in the following integral form with the use of {\boldmath${\cal H}_4$}:
\begin{eqnarray} 
  l_{ij}(T)  
  & = &   
  l_{ij}(T_r) 
  \: + \:
  \left( c_p^j - c_p^i \right)
  \left( T - T_r \right)
\; \: = \; \: 
  (l_{ij})_r 
  \: + \:
  \left( c_p^j - c_p^i \right)
  T_r \; X
\: , \label{eq10} \\
  R^1 \: T_r \: 
  \ln \left(\frac{p^{i1}(T)}{p^{i1}(T_r)} \right)
  & = &   
  l_{i1}(T) \:
  \left( 1 \: - \frac{T_r}{T}\: \right)
  \: - \:
  \left( c_p^1 - c_p^i \right) \:
   T_r \; {\cal F}(X)
\: , \label{eq11}
\end{eqnarray} 
where $T_r$ is a constant temperature. 
Equation (\ref{eq11}) holds for $ i = 2$ or $3$. 
The function ${\cal F}(X) = X - \ln( 1 + X)$ is 
introduced with $X = (T - T_r)/T_r$. 
\footnote{\color{blue} 
It is the same function ${\cal F}(X) = X - \ln( 1 + X)$
which is used to define the (flow-exergy) temperature component 
by $a_T = c_{pd} \: T_r \: {\cal F}(T/T_r-1)$  in  Marquet (1991).}

 \section{\underline{THE MOIST AVAILABLE ENTHALPY}.} 
\label{section_3}

Section \ref{section_3.1} deals with the dry and moist versions of the ``potential change in total entropy''. 
The results of the moist version are used in section \ref{section_3.2} as a starting point to define $a_m$. 
However, section \ref{section_3.1} may well be omitted on a first reading of this paper, considering section \ref{section_3.2} as the mathematical definition of the local function $a_m$.

         \subsection{\Large Introduction.} 
        \label{section_3.1}

The concept of dry ``potential change in total entropy'' is defined in MM. 
It is denoted by $\Delta_d \, S^o = a_h /T_r$ and can be compared with 
the same expression $\Delta \, S_t = R_{\rm max} /T_r$ of Landau and Lifchitz (1976, $\S$20).

\begin{figure}[t]
\centering
\includegraphics[width=0.49\linewidth,angle=0,clip=true]{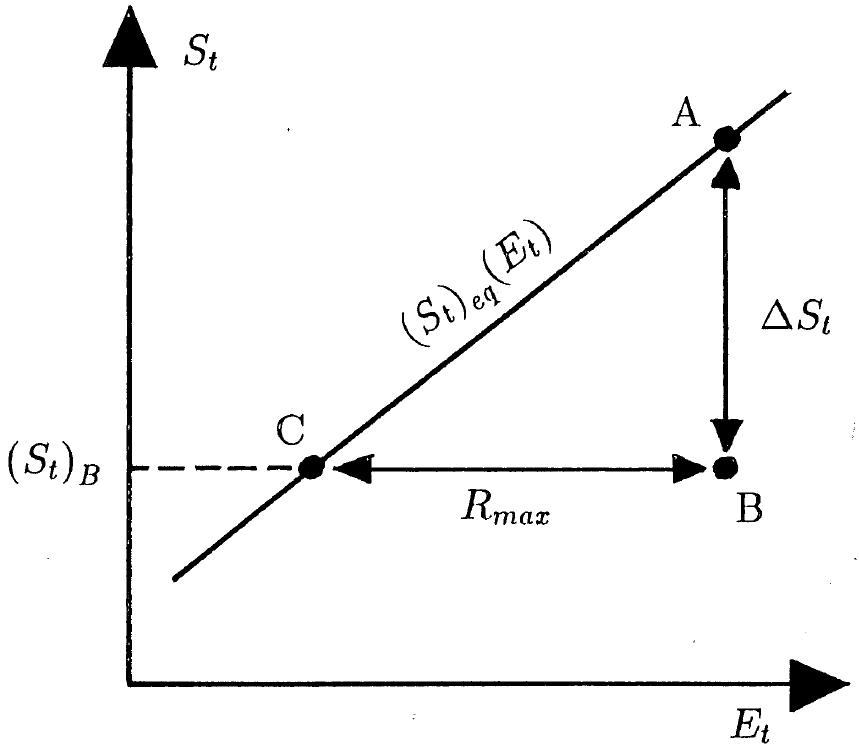}
\caption{\it
$S_t$--$E_t$ diagram diagram with the equilibrium curve
$(S_t)_{\rm eq}$ 
and the non-equilibrium state B. 
$\Delta \, S_t$ is the change in total entropy and 
$R_{\rm max}$ is the available energy.
\label{Fig_1}}
\end{figure}

Landau and Lifchitz have used a geometric approach following the method introduced
by Gibbs (1873).\footnote{\color{blue} 
The ``total entropy'' approach is described with some 
details in the arXiv version of Marquet (1991): 
\url{http://arxiv.org/abs/1402.4610} 
{\tt arXiv:1402.4610 [ao-ph]}.}
The problem is to define the available part of the total energy,
namely the part of the energy that a system can release when undergoing a spontaneous
process towards an equilibrium state. The system consists of a body surrounded by a
medium at constant pressure $p_0$ and temperature $T_0$. The total entropy of the body and
medium is $S_t$ and the total energy is $E_t$. 
Use of an $S_t$--$E_t$ diagram enables each state of
the system to be represented by a point. 
The equilibrium curve $(S_t)_{\rm eq}$ is the so-called
``surface of dissipated energy'' introduced by Gibbs (1873); 
it only depends on $E_t$ and is
a straight line if the medium is a thermostat (see Fig.~\ref{Fig_1}). 
The slope of this equilibrium curve is (approximately \footnote{\color{blue} 
The approximation comes from the derivative 
${d(S_t)_{\rm eq}}/{dE_t}$ which is 
approximated by the finite differences ${\Delta S_t}/
{\Delta E_t}$, and if the equilibrium 
curve were not a straight line (i.e. if $T_0$ were not 
a constant and would vary with $E_t$).})
related to the equilibrium temperature by
\begin{equation}
   \boxed{\;
\frac{1}{T_0}
\; \: \equiv \; \: 
\frac{d(S_t)_{\rm eq}}{dE_t} 
   \; }
\; \: \approx \; \: 
\frac{\Delta S_t}{\Delta E_t} 
\; \: = \; \: 
\frac{\Delta S_t}{R_{\rm max}} 
\; \;\: \Longrightarrow \; \: 
   \boxed{\;
   \Delta S_t
      \;  \approx \; 
      \frac{R_{\rm max}}{T_0}
   \; } \: .
\nonumber
\end{equation}
If the body is not in equilibrium with the medium and if the 
point B represents this general state in Fig.~\ref{Fig_1},
the distance BA is the ``capacity for entropy'' $\Delta \, S_t$, 
whereas the distance CB is the ``available energy'' $R_{\rm max}$
(this terminology was used by Gibbs), or the maximum available work.
According to the second law of thermodynamics any real 
thermodynamic process results in an increase of the total entropy. 
Therefore, starting from B, the attainable equilibrium points are located above
the value $(S_t)_B$, 
and the maximum variation in total energy is reached when the equilibrium
end state is C. 
This special process is the reversible one where the final total entropy has
not changed $(S_t)_B = (S_t)_C$. 
The consequence of the second law is that the distance CB is
a maximum and represents the available energy. 
The change in total entropy and the available energy 
are thus simply connected by the slope of the equilibrium curve: 
$\Delta S_t \approx R_{\rm max} / T_0 $.

The available enthalpy $a_h$ can also be related to a change 
in total entropy $\Delta_d \, S^o$:
\begin{equation}
{\Delta_d S^o}
\; \: = \; \: 
\frac{a_h}{T_r} 
\; \: = \; \: 
\frac{h(T,p) - h_r(T_r, p_r)}{T_r} 
\; - \;
\left[\;
 s(T,p)  - s_r(T_r, p_r)
\; \right]
\: ,
\label{eq12}
\end{equation}
where two undetermined constants $T_r$ and $p_r$ are introduced. 
It is demonstrated in Marquet (1991) that, for the dry atmosphere case,
${\Delta_d S^o}$  is the change in total entropy of a conceptual system 
undergoing two prescribed processes. 
This system consists of a unit mass parcel of dry ideal gas at $(T, p )$, 
and a mere conceptual thermostat at $T_r$.
The dry irreversible processes are: 
\begin{itemize}[label=,leftmargin=3mm,parsep=0cm,itemsep=0.1cm,topsep=0cm,rightmargin=2mm]
\vspace*{-1mm}
\item  i) first an isobaric change of the parcel from $(T, p)$ to $(T_r, p)$
coming into contact with the thermostat;
\item ii) then a sudden isothermal and adiabatic change from 
$(T_r, p)$ to $(T_r, p_r)$. 
\end{itemize}
In fact the second process involves a work reservoir at $p_r$, but the change
in total entropy ${\Delta_d S^o}$ is only related to the unit mass parcel and 
the thermostat.
The work exchanged between the parcel and the work reservoir is not considered as
available for the system, the work reservoir is a free source of work. This is the main
difference between the concept of available energy and available enthalpy.

For the available energy case one can similarly consider a unit mass parcel associated
with a work and heat reservoir at $T_r$ and $p_r$. 
The two new irreversible processes are: 
\begin{itemize}[label=,leftmargin=3mm,parsep=0cm,itemsep=0.1cm,topsep=0cm,rightmargin=2mm]
\vspace*{-1mm}
\item i) first a sudden and adiabatic change from $( T , p )$ to $( T_1 , p_r )$ coming into contact
             with the work reservoir, where $T_1$  is the temporary equilibrium value; and
\item ii) then an isobaric change from $( T_1 , p_r )$  to $( T_r , p_r )$  coming into contact 
            with the thermostat. 
\end{itemize}
The changes in total entropy $\Delta_1 \, S_t$ and $\Delta_2 \, S_t$ 
occurring during these two new processes lead to the following available
energy formulation: 
\begin{equation}
a_e 
\; \: = \; \: 
T_r \;
\left( \: \Delta_1 \, S_t + \Delta_2 \, S_t \: \right) 
\; \: = \; \: 
 \left[ \; e_i - (e_i)_r \: \right]
 \: + \: 
   p_r \: 
  \left( \frac{1}{\rho} - \frac{1}{\rho_r} \right) 
  \: - \:
  T_r \: \left(s - s_r \right)
\; . \nonumber
\end{equation}
This is the available energy function used among others by Landau and
 Lifchitz (1976), Livezey and Dutton (1976) and Karlsson (1990).

The moist-air generalization that is considered in this paper is made of:
\begin{itemize}[label=,leftmargin=3mm,parsep=0cm,itemsep=0.1cm,topsep=0cm,rightmargin=2mm]
\vspace*{-1mm}
\item 1) 
a unit mass moist-air parcel verifying {\boldmath${\cal H}_{1, ..., 9}$}  assumptions;
\item 2) 
a conceptual thermostat at $T_r <  T_{tr} \approx  273.16$~K;
\item 3) 
a moist-air {\it dead state\/}  at temperature $T_r$ and total pressure $p_r$;
\item 4) 
and finally four thermochemical processes. 
\end{itemize}
The moist-air potential change in total entropy is denoted by ${\Delta_m S^0}$.

The {\it dead state\/} is a saturated moist-air parcel at $T_r$, in possible chemical equilibrium
with an infinite and plane surface of ice, thus $p_r^1= p^{31}(T_r)$ according to
{\boldmath${\cal H}_{5}$}. 
There is no condensed phase: $r^2 = r^3 = 0$ and $r^1= r^{31}$
(it is the saturation mixing ratio over ice). 
The total pressure $p_r  = p^0_r + p^1_r$ and the temperature 
$T_r$ are supposed to be two constants. 
From {\boldmath${\cal H}_{5}$} the partial pressure of the 
dry-air component in the dead state only depends on $T_r$
and $p_r$:   $p^0_r (T_r, p_r) = p_r - p^{31}(T_r)$.

The {\it dead state\/} is a passive environment in that the chemical potential of water
vapour, $\mu_r^1$, and the chemical potential of ice, $\mu_r^3$, are the same (ice is the stable condensed
phase at $T_r\leq 0\,{}^{\circ}$~C). 
The partial chemical potential is defined in thermodynamics as 
$\mu^k (T, p) = h^k( T ) - T \: s^k ( T, p)$, for each component $k = 0$ to $3$.

Let us define the following sequence of thermochemical processes undergone by the
unit mass moist parcel and involving the thermostat:
\begin{itemize}[label=,leftmargin=3mm,parsep=0cm,itemsep=0.1cm,topsep=0cm,rightmargin=2mm]
\vspace*{-1mm}
\item 
(i) an isothermal ($T$) and adiabatic reversible separation of the parcel into its different
parts. The separation of the gases is realized via semi-permeable membranes. The
condensed phases are in equilibrium with the respective gas partial pressures. The
final state can be called the separate component state. This can be summarized by:\\
( $p^0$ , $p^1$ ; $m^k$ , $k = 0$ to $3$ ) 
$\; \to \;$ 
( $p^0$, $m^0$)  $\: + \:$ ( $p^1$, $m^1$)  $\: + \:$
( $p^{21}$, $m^2$)  $\: + \:$ ( $p^{31}$, $m^3$) ;
\item
(ii) if metastable phases are present (e.g. supercooled water) they are transformed
(at $T$ and $p$) into the more stable form at $T$, using heat exchanges with the thermostat
if needed;
\item
(iii) an irreversible and isobaric change from $T$ to $T_r$ coming into contact with the
thermostat, the final phase being the more stable one at $T_r$; and
\item
(iv) a sudden and isothermal irreversible change of pressure at $T_r$, to reach
( $p^0_r$, $T_r$) conditions for dry air and ( $p^{31}_r$ , $T_r$)
for all components $1$ to $3$ (the final phase
being the more stable one at $T_r$). 
The work reservoir involved in this adiabatic
process is supposed to be a free source of work.
\end{itemize}

The change in total entropy is zero for all components during the adiabatic and
irreversible process (i). 
Introducing the notation 
$b_k = (h^k - h^k_r ) - T_r \: (s^k - s^k_r)$
where the subscript $r$ denotes the (reference) dead state, 
the value of the moist change in total entropy for the
dry-air component $(\Delta_m S^o)^0$ 
along the path (i) to (iv) can be computed according to
Marquet (1991): 
$(\Delta_m S^o)^0 = b^0/T_r$. 
Indeed, (iii) and (iv) correspond in this case to the
dry processes of MM recalled at the beginning of this section, and the processes (i) and
(ii) are ineffective for the dry-air component.

\begin{figure}[t]
\centering
\includegraphics[width=0.95\linewidth,angle=0,clip=true]{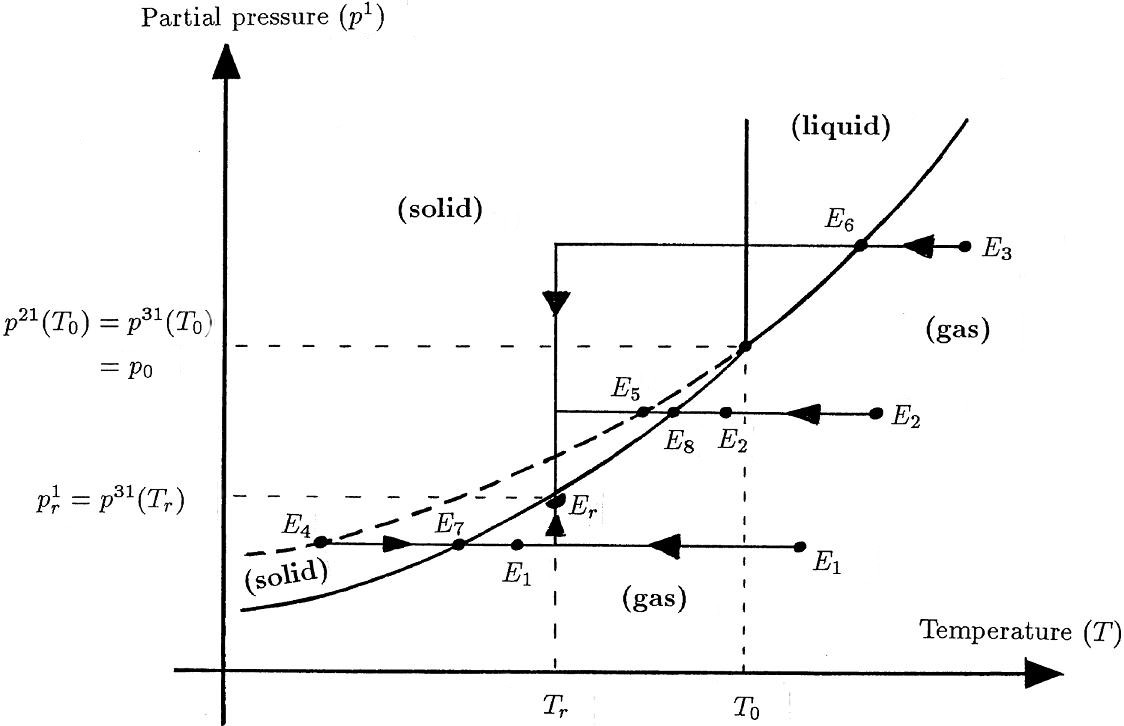}
\caption{\it
p-T diagram with a representation of the different paths for each water component between the
real states $E_{1, \ldots, 8}$ and the fixed thermodynamic reference state $E_r$.
\label{Fig_2}}
\end{figure}

It is more difficult to carry out the computation of $(\Delta_m S^o)^k$  
for the water components
($k = 1$, $2$, $3$) because numerous cases must be investigated depending on the various
changes of phase that can occur along the path (i) to (iv). There are in fact eight different
cases. Each of them is depicted in the $p$-$T$ diagram of Fig.~\ref{Fig_2} by possible initial points
$E_{1, \ldots, 8}$ and the common final state $E_r$ (the gaseous saturated dead state). 
If $p^1_r = p^{31}_r (T_r)$ and 
$p_0 = p^{31} (T_0) = p^{21} (T_0)$, 
where $T_0= T_{\rm tr}$, 
these states can be briefly defined as:
\begin{itemize}[label=$\bullet$,leftmargin=3mm,parsep=0cm,itemsep=0.1cm,topsep=0cm,rightmargin=2mm]
\vspace*{-1mm}
\item 
$E_1$: water vapour at $p^1 \leq p^1_r$ (always gaseous)
\item 
$E_2$: water vapour at $p^1$ and with $p^1_r \leq  p^1 \leq p_0$  (gas $\: \to \:$ solid $\: \to \:$ gas)
\item 
$E_3$: water vapour at $p^1_r \geq  p_0$  (gas $\: \to \:$  liquid $\: \to \:$ solid $\: \to \:$  gas)
\item 
$E_4$: supercooled liquid water at $T \leq T_r$ (liquid  $\: \to \:$   solid  $\: \to \:$   gas)
\item 
$E_5$: supercooled liquid water at $T$ and with $T_r \leq T  \leq T_0$ (liquid $\: \to \:$   solid $\: \to \:$   gas)
\item 
$E_6$: normal liquid water at $T \geq T_0$ (liquid $\: \to \:$ solid $\: \to \:$  gas)
\item 
$E_7$: ice at $T \leq T_r$ (solid $\: \to \:$  gas)
\item 
$E_8$: ice at $T$ and with $T_r \leq T  \leq T_0$  (solid $\: \to \:$ gas).
\end{itemize}

The computations of $(\Delta_m S^o)^k$ are somewhat long, 
but they are consistent since the cases $E_{1, 2, 3}$, 
$E_{4, 5, 6}$  and $E_{7, 8}$  respectively lead to the 
following expressions $a_m^1$, $a_m^2$ and
$a_m^3$  for the partial moist available enthalpies:
\begin{eqnarray}
a_m^0  & = & T_r \;  (\Delta_m S^o)^0
            \; \:  =  \; \:  b^0 \: ,
\label{eq13} \\
a_m^1  & = & T_r \;  (\Delta_m S^o)^1
            \; \:  =  \; \:  b^1  
                 \: + \: \left(  \mu_r^1 - \mu_r \right) 
            \; \:  =  \; \:  b^1 \: ,
\label{eq14} \\
a_m^2  & = & T_r \;  (\Delta_m S^o)^2
            \; \:  =  \; \:  b^2 
                 \: + \: \left(  \mu_r^2 - \mu_r \right) \: ,
\label{eq15}\\
a_m^3  & = & T_r \;  (\Delta_m S^o)^3
            \; \:  =  \; \:  b^3 
                 \: + \: \left(  \mu_r^3 - \mu_r \right)
            \; \:  =  \; \:  b^3 \: .
\label{eq16}
\end{eqnarray}

The result $a_m^0$  for $k = 0$ is recalled and 
$\mu_r = \mu_r^1 = \mu_r^3$
denotes the reference chemical
potential of water elements in the 
dead state ($\mu_r^k = h_r^k - T_r \: s_r^k$). 
The hypotheses ${\cal H}_{1,2,4,5,6,7}$  together with the integral forms
(\ref{eq10}) and (\ref{eq11}) of Eq.(\ref{eq09}) have all been used to work out
Eqs.(\ref{eq13}) to (\ref{eq16}).

         \subsection{\Large Definition of $a_m$.} 
        \label{section_3.2}

Let us assume that a, is an extensive state function $a_m = m^k \: a_m^k$. 
The subscript $r$ denotes the moist dead state described in section \ref{section_3.1}. 
It is a passive environment represented by a saturated moist parcel at 
$T_r \leq T_0 = 273.16$~K and without a condensed phase. 
The partial available enthalpies $a_m^k$ are defined by 
Eqs.(\ref{eq13}) to (\ref{eq16}) where 
$b^k = (h^k - h_r^k) - T_r \: (s^k - s^k_r)$.
Using these definitions the analytical expression of the local 
moist available enthalpy is
\begin{eqnarray}
a_m  & = & 
    m^k \: (  h^k - h_r^k )
   \; - \;
   T_r \; m^k \: (  s^k - s_r^k )
   \; + \;
   m^2 \: (  \mu^2_r - \mu_r^3 )
 \: .
\label{eq17}
\end{eqnarray}

At first sight one could note an asymmetry, with an arbitrary importance given to liquid
water through $m^2$; but this equation really is symmetric with respect to water components
($i = 1$ to $3$). The two terms $m^1 \: ( \mu^1_r - \mu^3_r)$ and 
$m^3 \: ( \mu^3_r - \mu^3_r)$ coming from Eqs.(\ref{eq14}) and
(\ref{eq16}) could be included in Eq.(\ref{eq17}), 
but they are zero because $\mu^1_r = \mu^3_r$. The only
asymmetry is the fact that ice is a more stable component 
than liquid water at $T_r$.

An alternative form can be derived using Eqs.(\ref{eq05}) and 
(\ref{eq06}) with $\mu^k_r = h^k_r - T_r \:s^k_r$ to give
\begin{eqnarray}
a_m  & = & 
    h - T_r \; s 
   \; - \;
   m^k \: \mu_r^k 
   \; + \;
   m^2 \: (  \mu^2_r - \mu_r^3 )
 \: .
\nonumber
\end{eqnarray}
where the last term $\mu_r^3$  can be replaced by $\mu_r^1$. 
The transformation (\ref{eq04}) is then applied to the implicit sum
$- m^k \: \mu_r^k$,  with $\mu_r^3 =\mu_r^1$, 
yielding\footnote{\color{blue}
Eq.~(\ref{eq18}) can be written as 
$\boxed{\: a_m   =   \: h - T_r \; s 
   \: - \:  q_d \: (\mu_r)_d 
   \: - \:  q_t \: (\mu_r)_v }\;$.
These boxed formulae is important because they are expressed in terms 
of the basic moist-air enthalpy ($h$) and entropy ($s$)
of any parcel of atmosphere.
The last two terms depends on the dry-air and total-water 
contents, namely $q_d$ and $q_t$.
Moreover, the last two terms can be transformed 
by using $q_d = 1 - q_t$, leading to
$\boxed{\: a_m   =   \: h - T_r \; s  
  \: - \:  q_t \: [ \:  (\mu_r)_v - (\mu_r)_d \: ] 
  \: - \: (\mu_r)_d  }\;$.
These formulae are clearly symmetric with respect to water species, 
simply because $q_t = q_v + q_l + q_i$ is invariant 
with respect to changes of phases like $q_v \leftrightarrow q_l$
 which corresponds to condensation or evaporation processes.
Moreover, for closed parcel of fluid $q_t$ and $q_d$ are constant 
terms and therefore the available enthalpy $a_m$ varies like 
$( h - T_r \: s )$, up to true constant terms.
}
\begin{equation}
\boxed{\;
a_m  \; \:  = \; \: 
    h - T_r \; s 
   \; - \;
   m^0 \: \mu_r^0 
   \; - \;
   m^t \: \mu_r^1
  \; }
 \: .
\label{eq18}
\end{equation}

From Eq.~(\ref{eq17}) one can verify that $a_m$ is 
independent of any arbitrary choice for the
absolute values of energy, enthalpy or entropy. 
Indeed the difference in enthalpies $h^k - h_r^k$ 
and in entropies $s^k - s_r^k$  can be expressed
 in terms of $T - T_r$ and $p^k / p_r^k$. 
They are respectively equal to
$c_p^k \; ( T - T_r)$ and 
$c_p^k  \: \ln(T /T_r) - R^k \: \ln(p^k / p_r^k)$, 
except for the entropy of the condensed
phases which is $c^k  \: \ln(T /T_r) $, 
$k = 2$ and $3$. 
The last term $(  \mu^2_r - \mu_r^3 )$ 
in Eq.(\ref{eq17}) corresponds
to the chemical affinities related to 
the water vapour: 
${\cal A}^i  
= \mu^1 - \mu^i 
= R^1 \: T \: \ln(p^1/p^{i1})$,
$i = 2$ or $3$. 
And ${\cal A}^2_r -  {\cal A}^3_r$  yields 
$\mu^2_r - \mu_r^3 
  = R^1 \: T_r \: \ln[ \: p^{21}(T_r)/p^{31}(T_r) \: ]$,
which depends on saturation pressures which only depend 
on the temperature $T_r$.

On the contrary, the values of zero-entropies are used in HH to derive the entropy
temperature. Standard chemical values for enthalpy and entropy are also introduced in
many technical papers in order to derive the ``standard chemical exergy'' for each chemical
element (see, for example, Szargut (1980); Ahrendts (1980); Morris and Szargut (1986)).

Equation (\ref{eq18}) gives the more general definition 
of $a_m$ in terms of $h$ and $s$. 
It must be remarked that each term of this expression 
(i.e. $h$, $s$, $\mu_r^0$ and $\mu_r^1$) cannot be separately
determined without assuming unnecessary agreed reference 
values for $h$ and $s$. 
Indeed, even if $h^1$, $h^2$ and $h^3$ are 
linked through the latent heats $l_{ij} = h^j - h^i$, 
$h^0$ cannot be connected with the others for 
lack of chemical reaction between dry air and water
elements. 
Nevertheless, Eq.(\ref{eq18}) is interesting in that it reveals 
that $a_m$ exactly varies as
$h - T_r \: s$ for a closed parcel (i.e. if $m^0$ and 
$m^t$ are two constants). Moreover $a_m$ simply
varies as $h$ for a closed parcel with adiabatic 
and reversible motion ($s$ is then also a constant
in that case). This last property will be investigated 
in section \ref{section_8}, yielding the definition
of a conservative potential temperature.

The moist available enthalpy defined by Eqs.(\ref{eq17}) 
and (\ref{eq18}) is different from the flow
essergy function $e_i + p_0 \ v - T_0 \: s - m^k \: \mu^k_0$  
(Evans 1969 or 1980; Landau and Lifchitz 1976
section~96; Livezey and Dutton 1976; Karlsson 1990). 
The local approach of Karlsson and the
global approach of Livezey and Dutton 
are different from the present one because in
their papers the reference state is defined with 
a reference pressure $p_0^k( z )$ which varies
as $\exp(-z/H^k)$. 
The reference entropy $s_0^k  [ \: T_0 , \: p_0^k ( z ) \:  ]$
is not a constant, and the reference
chemical potential $\mu_0^k ( z )$ 
is equal to $h_0^k - T_0 \:  s_0^k (z)$.
Equation~(\ref{eq18}) is more similar to the
definition of the stationary flow availability 
function $h - T_0 \: s_0^k - m^k \mu_0^k$ 
given by Evans and quoted in Ahrendts (1980), 
or the flow exergy $h - T_0 \: s - m^k \mu_0^k (z)$ 
defined in Karlsson
(1990), except for the term $m^t \mu_r^1$  
in place of $\sum_{k=1}^{3} m_k \: \mu_0^k$.

         \subsection{\Large The components of $a_m$.} 
        \label{section_3.3}

In order to show in what sense $a_m$ is a generalization of the dry function $a_h$ (MM),
$a_m$ must be further transformed. Starting from Eq.(\ref{eq17}) the first two terms 
$m^k \: (h^k - h_r^k) = c_p \:  (T - T_r)$ and $m^k \: T_r \: (s^k - s_r^k)$ 
can be transformed with Eq.(\ref{eq04}) to give respectively
\begin{equation}
m^k \: (h^k - h_r^k) 
\; \: = \; \:
\left[ \: 
m^0 \: c_p^{\ast}  
\: - \:  m^2 \: ( c_p^{1}  - c^2 )
\: - \:  m^3 \: ( c_p^{1}  - c^3 )
\: \right] \: T_r \: X
 \: ,
\nonumber
\end{equation}
and
\begin{eqnarray}
m^k \: T_r \: (s^k - s_r^k)
& = &
\left[ \: 
m^0 \: c_p^{\ast}  
\: - \:  m^2 \: ( c_p^{1}  - c^2 )
\: - \:  m^3 \: ( c_p^{1}  - c^3 )
\: \right] \: T_r \:\ln(1+ X)
\nonumber \\
& &
\: - \: m^0 \: R^0 \:  T_r \: 
       \ln\!\left(  \frac{p^0}{p^0_r} \right)
\: - \: m^t \: R^1 \:  T_r \: 
       \ln\!\left(  \frac{p^1}{p^1_r} \right)
\nonumber \: .
\end{eqnarray}

The specific heats $c_p = m^k \: c_p^k$ 
and $c_p^{\ast} = c_p^0 + r^t \:  c_p^1$ 
have been introduced as well as the
variable $X= ( T - T_r)/T_r$. 
The last term of Eq.(\ref{eq17}) 
is computed as 
$\mu_r^2 - \mu_r^3 = R^1 \: T_r \: \ln(p_r^{21}/p_r^{31})$. 
A new form is then obtained for $a_m$ using the function 
${\cal F}(X) = X - \ln(1 + X)$:
\begin{eqnarray}
a_m  & = &
m^0 \: c_p^{\ast}  \;\: T_r \: {\cal F}(X) 
\: - \:  \left[ \: 
           m^2 \: ( c_p^{1}  - c^2 )
           \: + \: 
           m^3 \: ( c_p^{1}  - c^3 )
          \: \right]  T_r \: {\cal F}(X)
\nonumber \\
& &
\: + \: m^0 \: R^0 \:  T_r \: 
       \ln\!\left(  \frac{p^0}{p^0_r} \right)
\: + \: m^t \: R^1 \:  T_r \: 
       \ln\!\left(  \frac{p^1}{p^1_r} \right)
\: + \: m^2 \: R^1 \:  T_r \: 
       \ln\!\left(  \frac{p_r^{21}}{p_r^{31}} \right)
\nonumber \: .
\end{eqnarray}
The last step is to use the property $p_r^{31} = p_r^{1}$ 
together with Eq.(\ref{eq11}) to transform the last
term and the terms between square brackets 
into an expression involving pressures and
latent heats
\begin{eqnarray}
  &  &
 - \: ( \, m^2 \: l_{21} \: + \:  m^3 \: l_{31} \,  )
  \left(  1 \: - \: \frac{T_r}{T} \right)
 \; + \; 
 ( \, m^2 + m^3 \, ) \: R^1 \:  T_r \: 
       \ln\!\left(  \frac{p^1}{p^{31}_r} \right)
\nonumber \\
& &
\: - \: 
\left[ \;
       m^2 \: R^1 \:  T_r \: 
       \ln\!\left(  \frac{p^{1}}{p^{21}} \right)
      \: + \: 
       m^3 \: R^1 \:  T_r \: 
       \ln\!\left(  \frac{p^{1}}{p^{31}} \right)
\; \right]
\nonumber \: 
\end{eqnarray}
where the last term between square brackets is zero according to Eq.(\ref{eq08}).

After these manipulations one obtains the separation of $a_m$ into three moist energy
components:
\begin{eqnarray}
  a_m   &  =   &   a_T \; + \;  a_l \; + \;  a_p
\label{eq19} \: , 
\end{eqnarray}
where the temperature component is
\begin{eqnarray}
  a_T   &  =   &   m^0 \; c_p^{\ast} \; T_r \; {\cal F}(X) \; \geq \; 0
\label{eq20} \: , \\
  \mbox{because}  \;\;\;\; {\cal F}(X)    &  \geq   &    0
\label{eq21} \: , 
\end{eqnarray}
and where the latent (condensed-water) and pressure components are
\begin{eqnarray}
  a_l   &  =   &    - \: ( \, m^2 \: l_{21} \: + \:  m^3 \: l_{31} \,  )
                          \left(  1 \: - \: \frac{T_r}{T} \right)
\label{eq22} \: ,  \\
  a_p   &  =   &   
\: m^0 \: R^0 \:  T_r \: 
       \ln\!\left(  \frac{p^0}{p^0_r} \right)
\: + \: m^t \: R^1 \:  T_r \: 
       \ln\!\left(  \frac{p^1}{p^{31}_r} \right)
\label{eq23} \: .  
\end{eqnarray}

The dry case is defined by 
$m^1 = m^2 = m^3 = m^t = 0$, $m^0 = 1$, $p^0 = p$ and $p^0_r = P_r$, 
where $P_r$ is the reference pressure introduced in the dry available 
enthalpy approach (see the beginning of section \ref{section_3.1}. 
In the dry case $a_m$ reduces to $a_h = a_m^0 = b^0$, 
and both $a_T$  and $a_p$ are the same as the temperature 
and pressure components already defined in MM
($c_p^{\ast} \rightarrow c_p^0$ in Eq.(\ref{eq20})). 
The functions $a_T$ and $a_p$ in Eqs.(\ref{eq20}) 
and (\ref{eq23}) are called the moist temperature 
and pressure components of $a_m$.

The ``{\it latent\/}'' component $a_l$ is a new one. 
It is different from zero only if there is a
condensed phase inside the parcel (i.e. $m^2 \neq 0$ or $m^3 \neq 0$). 
Since $ l_{21} > 0$ and $l_{31} > 0$, the sign of 
$a_l$  is the same as $- \:(T - T_r)$. 
One can recognize in Eq.(\ref{eq22}) the so-called efficiency
factor $\eta_T = 1 - T_r/T$ introduced in the available 
and static entropic energy theories and
also used in the dry available enthalpy approach. 
But in the present moist theory $\eta_T$  appears
directly in the formulation of the latent-energy component, 
whereas it was involved in the previous studies only 
at the stage of the prognostic equations for the
energy components. 
It should be noticed that, 
from Eqs.(\ref{eq20}) to (\ref{eq22}), 
$a_T$ and $a_l$ vanish
if $T= T_r \Leftrightarrow {\cal F}(X) = 0  \Leftrightarrow \eta_T  = 0$ .

 \section{\underline{NUMERICAL EVALUATIONS}.} 
\label{section_4}

The local available enthalpy function $a_m$ 
was separated in section~\ref{section_3.3}
into the three components $a_T$, $a_l$ and 
$a_p$. 
They are natural generalizations of the 
temperature and pressure components derived 
for the dry case, namely 
$(a_T)^0 = c_p^0 \: T_r \: {\cal F}(X)$ 
and 
$(a_p)^0 = R^0 \: T_r \: \ln (p^0/P_r)$. 
The latent heat part $a_l$ does not exist in the dry case.

\begin{figure}[t]
\centering
\includegraphics[width=0.95\linewidth,angle=0,clip=true,angle=0.0]{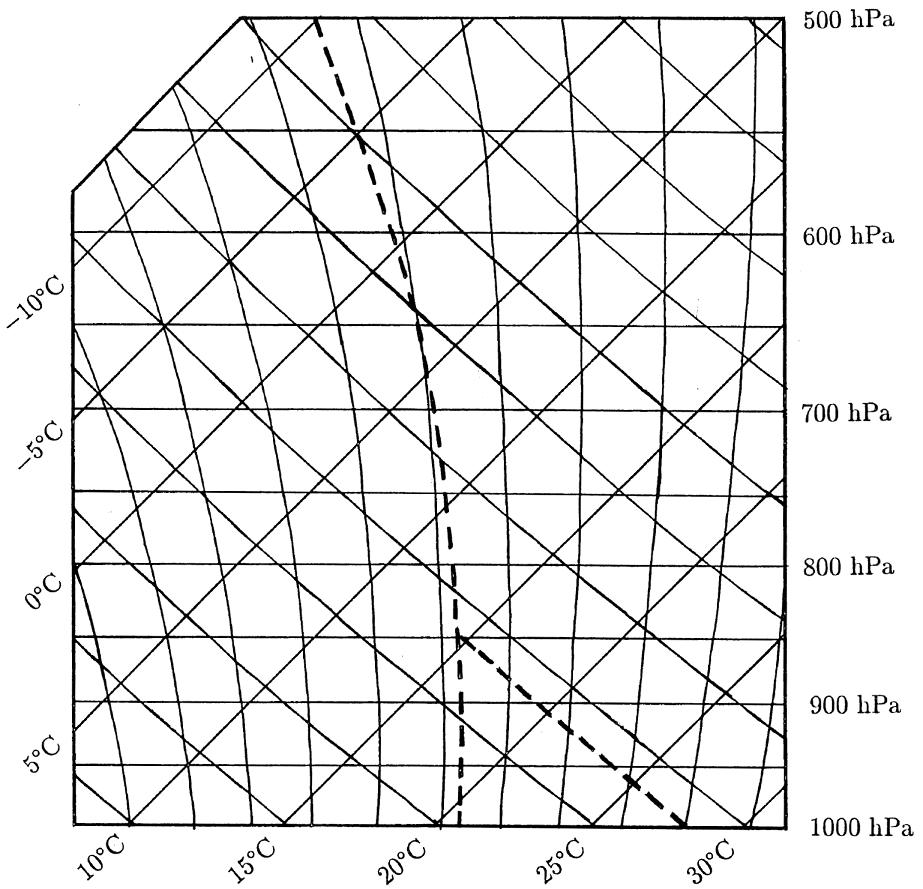}
\caption{\it
A representative vertical profile of a cumulus (dew points and state curve on a Skew-emagram for $p\geq 500$~hPa).
\label{Fig_3}}
\end{figure}

\begin{figure}[t]
\centering
\includegraphics[width=0.6\linewidth,angle=0,clip=true,angle=0]{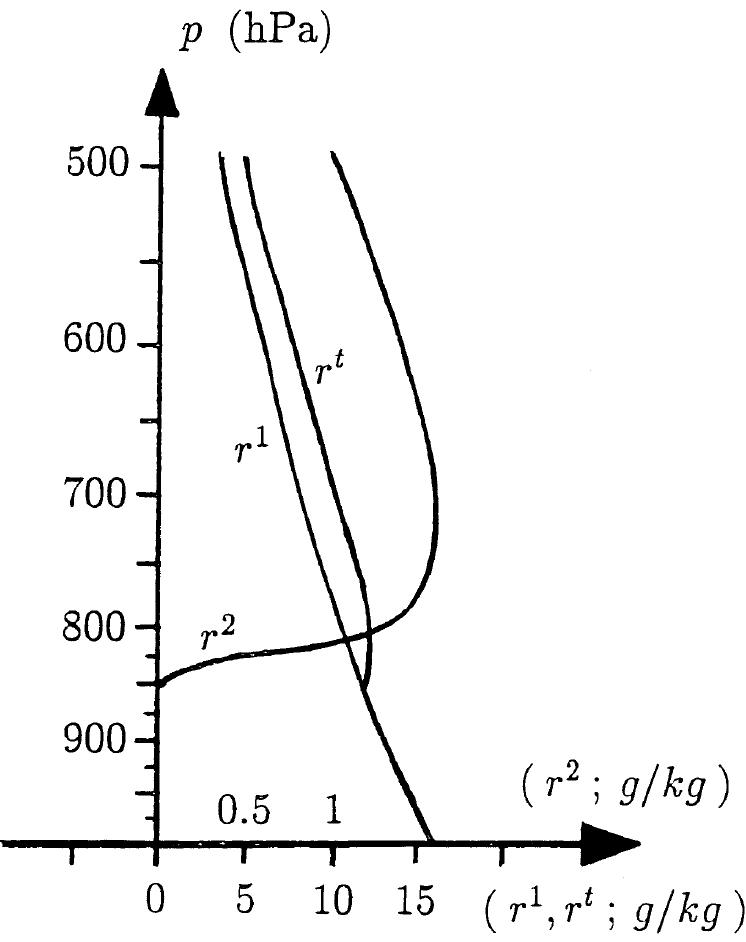}
\caption{\it
Profiles of the mixing ratios of water vapour, 
$r^1$, and cloud liquid water, $r^2$, with 
the corresponding profile of total mixing 
ratio of water, $r^t = r^1 + r^2$. 
Note that the scale for $r^2$ is
different from those for $r^1$ and $r^t$.
\label{Fig_4}}
\end{figure}

In this section an attempt is made to estimate the orders of magnitudes of these
moist components. A representative vertical profile of a cumulus cloud with liquid cloud
water is used (see Fig.~\ref{Fig_3}). 
There is neither ice nor precipitable liquid water. 
Air is saturated above the $850$~hPa level. 
The profiles of the mixing ratios of water vapour $r^1$, 
cloud liquid water $r^2$, as well as the total mixing ratio 
of water $r^t = r^1 + r^2$ are shown in
Fig.~\ref{Fig_4}. 
Values of $r^t$ and $r^1$ are about $10$~g~kg${}^{-1}$, 
and values of $r^2$ are about $1.5$~g~kg${}^{-1}$.

The impact of moisture on the definition of available enthalpy is firstly the appearance
of a new term $a_l$, secondly the modification of the dry components by terms depending
on $r^t$ ($a_T$ and $a_p$). 
The moist components can be rewritten as 
$a_T = m^0 \: \{ \: (a_T)^0 + (a_T)^t \: \}$ 
and
$a_p = m^0 \: \{ \: (a_p)^0 + (a_p)^t \: \}$,
with
\begin{align}
 (a_T)^0  & =  c^0_p  \;\: T_r \: {\cal F}(X) \; ,
 \hspace*{1cm}
 (a_T)^t  \; =  r^t \; c^1_p  \;\: T_r \: {\cal F}(X)  \; ,
\nonumber \\
 (a_p)^0  & =  R^0 \:  T_r \: 
       \ln\!\left(  \frac{p_r^0}{p^0_r} \right) \; ,
 \hspace*{0.5cm}
 (a_p)^t  \; =  r^t \; R^1 \:  T_r \: 
       \ln\!\left(  \frac{p_r^1}{p^1_r} \right)
\nonumber \: .
\end{align}
The (reference) dead state is defined by the constants $T_r = 251$~K, 
$p^1_r = 0.85$~hPa and $p^0_r = 369.15$~hPa.

\begin{figure}[t]
\centering
\includegraphics[width=0.6\linewidth,angle=0,clip=true,angle=0]{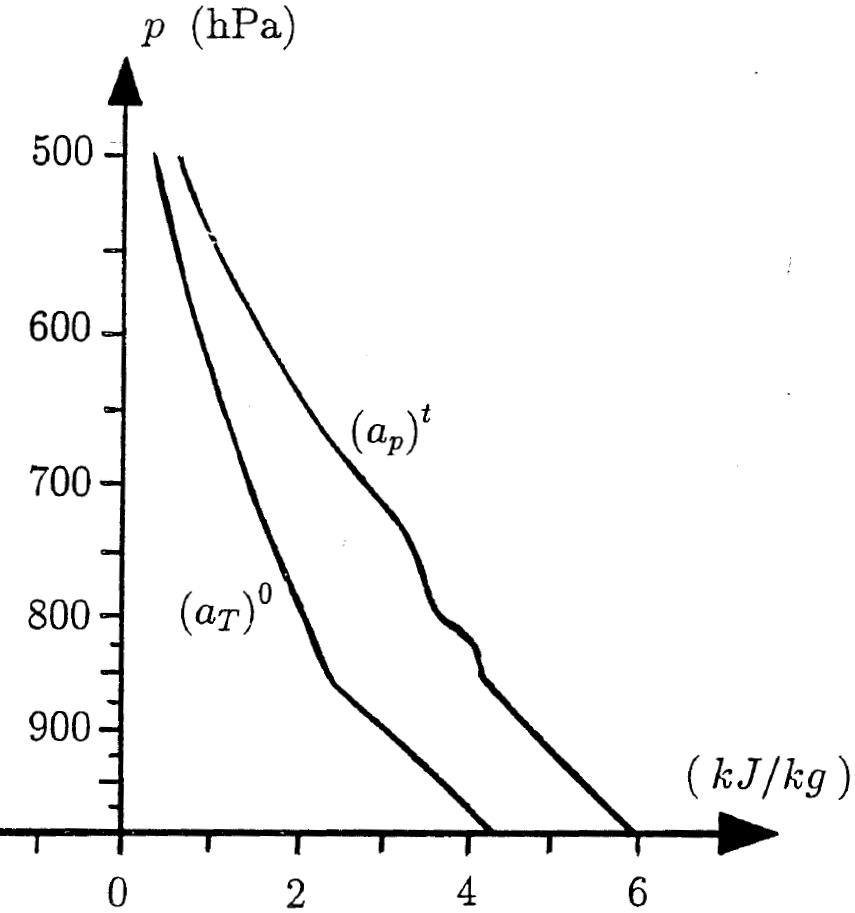}
\caption{\it
Vertical profiles of dry-air $(a_T)^0$ and water vapour $(a_p)^1$ components.
\label{Fig_5}}
\end{figure}

\begin{figure}[t]
\centering
\includegraphics[width=0.6\linewidth,angle=0,clip=true,angle=0]{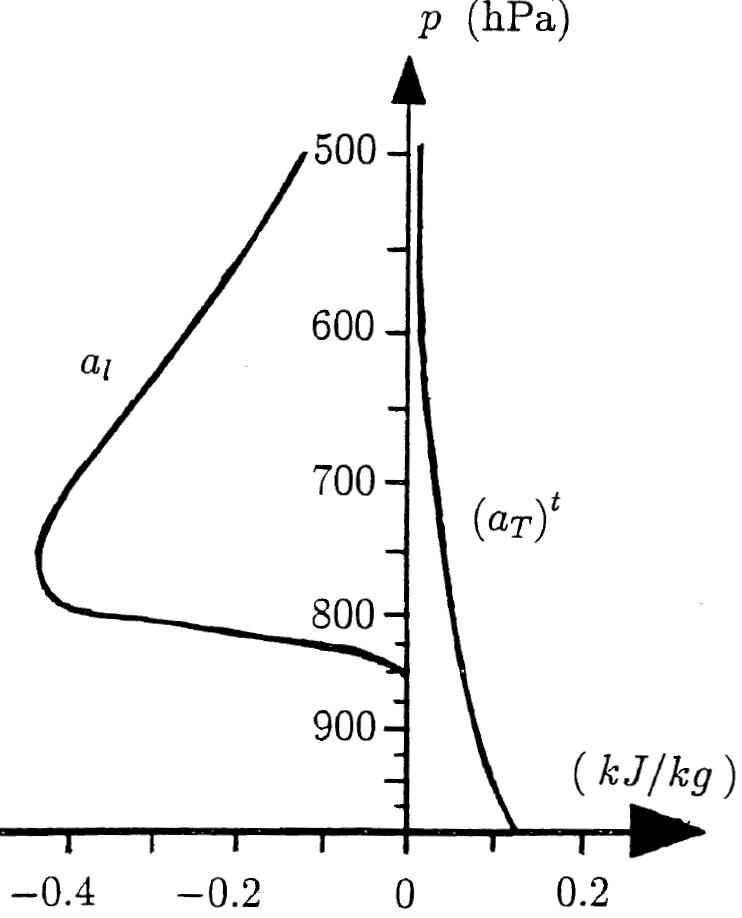}
\caption{\it
Vertical profiles of liquid-water $a_l$ and total-vapour $(a_T)^t$ components.
\label{Fig_6}}
\end{figure}

The vertical profiles of $(a_T)^0$  is depicted on Fig.~\ref{Fig_5}. 
It is similar to the lower part $p \geq  500 $~hPa, 
of Fig.~\ref{Fig_6}(a) in MM. 
The profile of the water pressure component is
also shown in Fig.~\ref{Fig_5}. 
The values of $(a_p)^t (p)$ are greater than the values of the dry-air
temperature component $(a_p)^0 (p)$. 
The moist correction to $a_p$  is thus an important one.
It can be as high as $5$~kJ~kg${}^{-1}$ 
and the ratio $(a_p)^t  / (a_p)^0$
(not shown) is about 7~\% for
$p \geq 700$~hPa.
  
The vertical profiles of $a_l$ and $(a_p)^t$ are depicted on Fig.~\ref{Fig_6}. 

The values of $a_l(p)$ are about $-0.4$~kJ~kg${}^{-1}$, 
one order of magnitude less than $(a_T)^0(p)$  or $(a_T)^t(p)$, 
but in fact it is an important contribution to the 
available enthalpy budget because $0.4$~kJ~kg${}^{-1}$
is a typical value for the dry baroclinicity component, 
see Fig.~\ref{Fig_6}(b) of Marquet (1991). 
This result still holds with the isobaric average 
of the latent component ($a_l$ is always negative
and the average value is close to the order of magnitude).

As for the values of $(a_T)^t(p)$ they do not exceed $0.1$~g~kg${}^{-1}$ -- which is small in
comparison with all other components --  but the ratio $(a_T)^t / (a_T)^0$ is greater than $2$~\% for
$p \geq 700$~hPa and it could be important to take $(a_T)^t $ into account  in a fine exergetic
budget.

These evaluations are not thorough however. The impacts of the ice cloud component
and the precipitable liquid water (or ice) should have been investigated. The precipitable
mixing ratio can indeed be as high as $8$~g~kg${}^{-1}$, 
a value which is five times the liquid cloud water content $r^2$. 
However, as explained in section~\ref{section_2}, liquid precipitation cannot be taken
into account in the present exergetic analysis (${\cal H}_1$). 
The effects of the presence of ice, either the cloud or the precipitable 
part, are not studied in the present paper. 
From ${\cal H}_1$ solid precipitation is indeed outside the scope of this 
exergy approach, although a profile of cloud ice crystals could have been considered.

The preliminary conclusion from these results is that the use of the moist available
enthalpy concept could prove to be useful in analyses of mesoscale systems where
conversions involving the terms $a_l$, $(a_T)^t$ and $(a_p)^t$  could be of the same magnitude as the
other main conversions.

 \section{\underline{THERMODYNAMIC EQUATIONS}.} 
\label{section_5}

To study the conversions that occur in local, mesoscale or global systems, the local
budget of $a_m$ must be related to the local budgets of the other forms of energy (potential,
kinetic). The objective of this section is thus to derive an analytical expression for
$d(a_m)/dt$, with $d/dt$ obtained from section~\ref{section_2} and 
$a_m$ from section~\ref{section_3.2} in terms of $h$ and $s$. 
The knowledge of the first (for $h$) and second (for $s$) laws of 
thermodynamics can thus serve this purpose.

Using ${\cal H}_{1,\ldots,9}$ and following De Groot and Mazur (1962), these laws can be rewritten
as
\begin{eqnarray} 
\frac{d\, h}{dt}  & = & 
                   \frac{1}{\rho} \:  \frac{d\, p}{dt}
           \: + \: ( \varepsilon  + q )
           \: - \: \frac{1}{\rho} \; {\bf J}^k  . \: {\bf \nabla} ( h^k )
           \: + \: h^k   \;  \frac{d_e}{dt}\, ( m^k )
\: , \label{eq24} \\
T \; \frac{d\, s}{dt} & = &  ( \varepsilon  + q )
           \: - \: \frac{1}{\rho} \; {\bf J}^k  . \: {\bf \nabla} ( h^k )
           \: + \: T \: s^k   \;  \frac{d_e}{dt}\, ( m^k )
           \: -  \: {\mu}^k   \;  \frac{d_i}{dt}\, ( m^k )
\: , \label{eq25}
\end{eqnarray} 
where the positive Rayleigh function $\varepsilon = (\rho)^{-1} \: ({\sigma} . {\bf \nabla}).{\bf v}$
 is the specific dissipation of
mechanical energy ($\boldmath{\sigma}$ is the viscous stress tensor), and where $q$ is the specific heating
rate: $q = - \: {\rho}^{-1} \:  {\bf \nabla} . \: {\bf J_{iq}}$. 
The inner heat flux, ${\bf J_{iq}}$, is caused by conduction and radiation, 
but not by changes of phase which are explicitly taken into 
account by the internal changes in enthalpy: 
$d_i (h^k)/dt$.

The prognostic equations for the barycentric velocity ${\bf v}$, the specific kinetic energy
of the barycentric motion ($e_K = {\bf v} . {\bf v} /2$) and the specific gravitational energy 
($e_G = \phi = g \: z$) are
\begin{eqnarray} 
\frac{d\, {\bf v}}{dt}  & = & 
            - \, \frac{1}{\rho} \:  \boldmath{\nabla} \, p
           \: + \, \frac{1}{\rho} \:  \boldmath{\nabla} . \, ( \: \sigma \: )
           \: + \: {\bf g}
           \: - \: f \: {\bf k} \times {\bf v}
           \: + \: {\bf F_r}
\: , \label{eq26} \\
\frac{d\, e_K}{dt} \; = \; 
             {\bf v} \: . \:  \frac{d\, {\bf v}}{dt}
            & = &  
           + \: {\bf g} \: . \:  {\bf v} 
           \:  - \, \frac{1}{\rho} \:   {\bf v} \: . \: \boldmath{\nabla} \, p
           \: - \: \varepsilon
          \: + \, \frac{1}{\rho} \:  \boldmath{\nabla} . \, ( \: \sigma \, . \:  {\bf v} \: )
          \:  + \: {\bf v} \: . \:  {\bf F_r} 
\: , \label{eq27} \\
\frac{d\, e_G}{dt} \; = \; 
             g \: \frac{d\, z}{dt}
            & = & 
           - \: {\bf g} \: . \:  {\bf v} 
\: . \label{eq28}
\end{eqnarray}
where ${\bf F_r}$, represents external forces like friction and where $f$ is the Coriolis parameter.

The local-budget equation for $a_m$, can then be obtained using Eq.~(\ref{eq18}) with
$D({\mu}^0_r)/Dt = D({\mu}^1_r)/Dt = 0$, leading to
\begin{align} 
\frac{da_m}{dt}  
  & \: = \; 
\frac{dh}{dt}  
\: - \:
T_r \: \frac{ds}{dt}  
\: - \:
{\mu}^0_r \: \frac{dm^0}{dt}  
\: - \:
{\mu}^1_r \: \frac{dm^t}{dt}
\: . \nonumber
\end{align}
With the use of Eqs.~(\ref{eq24}) and (\ref{eq25}), one obtains
\begin{align} 
\frac{da_m}{dt}  
 & \: = \; 
   \frac{1}{\rho} \:  \frac{dp}{dt}
   \: + \: \eta_T \: ( \varepsilon  + q )
   \: - \: \eta_T \: \frac{1}{\rho} \; {\bf J}^k  . \: {\bf \nabla} ( h^k )
  \nonumber \\
 & \quad 
   \: + \: (h^k - T_r \: s^k) \: \frac{d_e}{dt}\,( m^k )
   \: + \: \left(\frac{T_r}{T}\right) \, 
           \mu^k \; \frac{d_i}{dt}\,( m^k )
   \: - \: \left[ \:  
           \mu^0_r \: \frac{dm^0}{dt}
           \: + \:
           \mu^1_r \: \frac{dm^1}{dt}
           \: \right]
\: . \label{eq29}
\end{align}
\begin{itemize}[label=$\bullet$,leftmargin=3mm,parsep=0cm,itemsep=0.1cm,topsep=0cm,rightmargin=2mm]
\vspace*{-1mm}
\item 
The first term of the second line of Eq.~(\ref{eq29}) is the sum
$(h^k - T_r \: s^k) \: [ \: {dm^k}/{dt} - {d_i(m^k)}/{dt} \: ]$.
\item 
The second term is the sum
$(T_r/T) \: (h^k - T_r \: s^k) \: {d_i(m^k)}/{dt}$.
\item 
The terms between the square brackets can also be transformed 
using Eq.~(\ref{eq04}) applied to $\mu^k_r = h^k_r - T_r \: s^k_r$
to give the sum
$- \: (h^k - T_r \: s^k) \: {dm^k}/{dt}
 \: + \: (\mu^2_r - \mu^1_r) \: {dm^2}/{dt}$, where 
the property $\mu^3_r = \mu^1_r$ is used to cancel
the term containing ${dm^3}/{dt}$.
\end{itemize}
Adding these three sums leads to the budget equation for $a_m$.
\begin{align} 
\frac{da_m}{dt}  
 & \: = \; 
   \frac{1}{\rho} \:  \frac{\partial p}{\partial t}
   \: + \: \frac{1}{\rho} \:   {\bf v} \: . \: \boldmath{\nabla} \, p
   \: + \: \eta_T \: ( \varepsilon  + q )
   \: - \: \eta_T \: \frac{1}{\rho} \; {\bf J}^k  . \: {\bf \nabla} ( h^k )
  \nonumber \\
 & \quad 
   \: + \: (h^k - h^k_r) \: \frac{dm^k}{dt}
   \: - \: \eta_T \: h^k \: \frac{d_i}{dt}\,( m^k )
   \: + \: ( \mu^2_r \: - \: \mu^1_r ) \: \frac{dm^2}{dt}
\: . \label{eq30}
\end{align}
Similarly to Eq.~(\ref{eq17}) all the terms in Eq.~(\ref{eq30}) can be computed independently of any choice for the absolute value of $h$ or $s$.
This is true for 
${\bf \nabla} ( h^k )= c_p^k \: \nabla ( T )$, 
$h^k - h^k_r = c_p^k \: (T - T_r)$ 
and 
$ \mu^2_r \: - \: \mu^1_r = R^1 \: T_r \: \ln(p_r^{21}/p_r^{31})$. 
It is also true for the sum 
$h^k \: {d_i}( m^k )/{dt}$
which only depends on the latent heats 
``$l_{\alpha \, \beta}$'' 
and the chemical changes of $\alpha$ element into $\beta$ element 
``${d_i}( m^{\beta}_{\alpha} )/{dt}$'', leading to:
\begin{align} 
h^k \: \frac{d_i}{dt}\,( m^k )
 & \: = \; 
\frac{1}{2} \; l_{\alpha \, \beta} \; 
\frac{d_i}{dt}\,( m^{\beta}_{\alpha} )
\; = \;
l_{21} \; \frac{d_i}{dt}\,( m^1_2 )
\: + \:
l_{32} \; \frac{d_i}{dt}\,( m^2_3 )
\: + \:
l_{31} \; \frac{d_i}{dt}\,( m^1_3 )
 \: . \label{eq31}
\end{align}
where more precisely the changes in water vapor, liquid water and ice contents are
\begin{align} 
\frac{d_i}{dt}\,( m^1 )
 & \: = \; 
 + \: \frac{d_i}{dt}\,( m^1_2 ) \: + \: \frac{d_i}{dt}\,( m^1_3 )
 \: , \nonumber \\
\frac{d_i}{dt}\,( m^2 )
 & \: = \; 
 - \: \frac{d_i}{dt}\,( m^1_2 ) \: + \: \frac{d_i}{dt}\,( m^2_3 )
 \: , \label{eq32} \\
\frac{d_i}{dt}\,( m^3 )
 & \: = \; 
 - \: \frac{d_i}{dt}\,( m^2_3 ) \: - \: \frac{d_i}{dt}\,( m^1_3 )
 \: , \nonumber
\end{align}
respectively.

It must be remarked that Eq.~(\ref{eq30}) is a symmetric equation with respect to the water components $k = 1$ to $3$. 
The only asymmetry lies in the choice of a dead (reference) state saturated
with respect to ice ($\mu^3_r \: - \: \mu^1_r$). 
The consequence is the missing of the term 
$(\mu^3_r \: - \: \mu^1_r) \: \: {dm^3}/{dt}$.

Equation (\ref{eq24}) is the equation for the enthalpy; but neither enthalpy nor energy can be computed numerically for lack of knowledge of their absolute values, contrary to the temperature which is the physical associated concept and which is easy to measure. 
After some manipulations, using Eqs.~(\ref{eq01}) to (\ref{eq03}) and the property 
${\bf \nabla} ( h^k )= c_p^k \; \nabla ( T )$, an equation for the temperature can be derived from Eq.~(\ref{eq24}), yielding
\begin{eqnarray} 
c_p \; \frac{d\, T}{dt}  & = & 
                   \frac{1}{\rho} \:  \frac{d\, p}{dt}
           \: + \: ( \varepsilon  + q )
           \: - \: \frac{1}{\rho} \; c_p^k \; {\bf J}^k  . \: {\bf \nabla} ( T )
           \: - \: h^k   \;  \frac{d_i}{dt}\, ( m^k )
\: . \label{eq33}
\end{eqnarray} 
According to Eq.~(\ref{eq31}) the last term of this equation can be computed and, as expected, Eq.~(\ref{eq33}) is a well-founded prognostic equation for $T$.

 \section{\underline{LOCAL PROPERTIES}.} 
\label{section_6}

         \subsection{\Large A local energy cycle.} 
        \label{section_6.1}

The local energy cycle will not be investigated in its more general form. 
Some hypotheses will be made in order to symplify the formulae. 
The aim in this first study is to forget the contributions due to various irreversibilities.

After some manipulation the three terms on the second line of Eq.~(\ref{eq29}) can be put in the form
\begin{equation}
   \: + \: (h^k - T_r \: s^k) \: \frac{d_e \, m^k }{dt}
   \: + \: (\mu^0_r - \mu^1_r) \: \frac{d_e \, m^t}{dt}
   \: + \: \frac{T_r}{T} \; \left\{ \:  
           (\mu^2 \: - \: \mu^1 ) \: \frac{d_i \, m^2}{dt}
           \: + \:
           (\mu^3 \: - \: \mu^1 ) \: \frac{d_i \, m^3}{dt}
           \: \right\}
\: . \nonumber
\end{equation}
This last expression is obtained without approximation. 
Indeed, there is no chemical reaction between dry air and the water element, and from Eq.~(\ref{eq01}) $d_i \, m^0 / dt = d_i \, m^t / dt = 0$. 
Another consequence is 
$d \, m^0 / dt  = - d \, m^t / dt = - d_e \, m^t / dt$. 
Then the terms between square brackets of the second line of Eq.~(\ref{eq29}) can be rewritten as $(\mu^0_r - \mu^1_r) \: d_e \, m^t / dt$. 
As for the term 
$(T_r/T) \: \mu_k \: d_i \, m^k / dt$, 
it can be transformed using 
$d_i \, m^1 / dt = - \: d_i \, m^2 / dt - d_i \, m^3 / dt$.

The parcel is now supposed to be closed and inviscid 
$\sigma = 0, \varepsilon = 0, {\bf J}^k  = 0  
\Rightarrow \forall k: d_e \, m^k = 0$. 
The first and second terms of the last expression cancel because
$d_e \, m^k / dt = 0$. 
The last terms also cancel out as far as changes of phase are reversible (i.e. if $\mu$ is the same in each phase of the chemical reaction). 
Indeed, using Eqs.~(\ref{eq32}), these last terms can be rewritten as
\begin{equation}
    \frac{T_r}{T} \; \left\{ \:  
           (\mu^1 \: - \: \mu^2 ) \: \frac{d_i \, m^1_2}{dt}
           \: + \:
           (\mu^2 \: - \: \mu^3 ) \: \frac{d_i \, m^2_3}{dt}
           \: + \:
           (\mu^1 \: - \: \mu^3 ) \: \frac{d_i \, m^1_3}{dt}
    \: \right\}
\: \nonumber
\end{equation}
and if one of the $d_i \, m_{\beta}^{\alpha} / dt$ is not zero, the corresponding 
$(\mu^{\alpha} \: - \: \mu^{\beta})$ is zero. 
Therefore Eq.~(\ref{eq30}) reduces to the simplified form
\begin{equation}
    \frac{d a_m}{dt} \; = \;
    \frac{1}{\rho} \: \frac{d p}{dt}
           \: + \:
     \eta_T \: q
\: . \nonumber
\end{equation}

Under all these assumptions Eqs.~(\ref{eq27}), (\ref{eq28}) and that previous simplified form of Eq.~(\ref{eq30}) yield the local energy cycle for an inviscid and closed parcel:
\begin{eqnarray} 
\frac{d \, a_m}{dt}  & = & 
          \frac{1}{\rho} \:  \frac{\partial \, p}{\partial t}
     \: + \frac{1}{\rho} \:  {\bf v} \:  . \: {\bf \nabla} ( p )
     \: + \: \eta_T \: q
     \: , \nonumber \\
\frac{d \, e_K}{dt}  & = & 
        + \: {\bf g} \: . \: {\bf v}
     \: - \: \frac{1}{\rho} \:  {\bf v} \:  . \: {\bf \nabla} ( p )
     \: + \: {\bf v} \: . \: {\bf F_r}
     \: , \label{eq34} \\
\frac{d \, e_G}{dt}  & = & 
        - \: {\bf g} \: . \: {\bf v}
     \nonumber \: , 
\end{eqnarray} 
where the two conversion terms 
``$- \: {\rho}^{-1} \:  {\bf v}  \: . \: {\bf \nabla} ( p )$'' 
and ``$ {\bf g} \: . \: {\bf v}$'' 
appear with opposite signs on different equations.

The terms ``$- \: \varepsilon$'' and 
``${\rho}^{-1} \:  {\bf \nabla} \: . \: ( \sigma  \: . \: {\bf v} )$'' 
are zero in the kinetic equation according to the hypotheses of a closed and inviscid parcel. 
The local cycle (\ref{eq34}) is a moist generalization of the dry available enthalpy cycle investigated in MM and, like $a_h$, $a_m$ can be considered as a moist Lagrangian energy-like quantity.

If the previous assumptions were not verified, some other terms would appear in 
$d a_m / dt$ and in $d e_K / dt$. 
These terms would take account of irreversibilities caused by change of phase, diffusion fluxes or viscosity. 
Non-equilibrium thermodynamics deals with these irreversibility terms (see, for example, De Groot and Mazur 1962). 
A full local cycle with the five components $\{ e_G, e_K; a_T, a_p, a_l \}$ can also be obtained, separating $a_m$ into $\{ a_T, a_p, a_l \}$ in (\ref{eq34}).

         \subsection{\Large The Bernoulli equation.} 
        \label{section_6.2}

Adding the three equations of (\ref{eq34}) gives:
\begin{eqnarray} 
\frac{d}{dt} \left( a_m + e_K + e_G \right) & = & 
          \frac{1}{\rho} \:  \frac{\partial \, p}{\partial t}
     \: + \: \eta_T \: q
     \: + \: {\bf v} \: . \: {\bf F_r}
     \: . \label{eq35} 
\end{eqnarray} 
Equation (\ref{eq35}) shows that $a_m + e_K + e_G$ is a constant along any particular streamline of an adiabatic frictionless and reversible steady flow if the parcels are assumed to be inviscid and closed with, however, possible reversible internal changes of phase of water components inside them. 
Equation (\ref{eq35}) is the moist generalization of the dry Bernoulli
equation of Marquet (1991).

This moist Bernoulli equation demonstrates that, if the required assumptions are
verified, $a_m$, exactly varies as $- \: ( e_K + e_G)$  during the motion of the parcel. Therefore energy conversions {\it really occur\/} between the various forms $a_m$, $e_K$ and $e_G$.

According to the local cycle (\ref{eq34}) a set of conversion terms is given by
$C_{(G,K)} = - \: {\bf g} \: . \: {\bf v}$ 
and
$C_{(m,K)} = - \: {\rho}^{-1} \:  {\bf v}  \: . \: {\bf \nabla} ( p )$, 
insofar as the other terms are interpreted as sources and sinks.
As usual, however, this interpretation of physical conversions is not unambiguous
(Johnson and Downey 1082). Other sets of conversion or source/sink terms can be
defined in (\ref{eq34}) provided the Bernoulli equation (\ref{eq35}) remains unchanged.

 \section{\underline{INTEGRAL PROPERTIES}.} 
\label{section_7}

The objectives of this section are to derive a global conservation law verified by the
total moist available enthalpy $A_m$, and to give definitions for the two undetermined
quantities $T_r$ and $p^0_r$.

The local moist property 
\begin{equation}
\frac{1}{\rho} \: \frac{\partial \, p}{\partial t}
= \frac{d}{dt}(R \: T)
\: - \: \frac{1}{\rho} \: 
\: {\bf \nabla} \: . \: ( \, p \:  {\bf v})
\nonumber
\end{equation}
holds true, provided that 
$R = m^0 \: R^0 + m^t \: R^1$ and $p = p^0 + p^1$. 
Starting with the integral form of this moist property together with the integral form of Eq.~(\ref{eq35}), it can be demonstrated that for an adiabatic frictionless and reversible motion of inviscid and closed atmospheric parcels:
\begin{equation}
\frac{d}{dt}(A_m + E_K)
 \: = \:  \frac{d}{dt}\left[ \: H - (E_i \: + \: E_G) \: \right]
 \: . \label{eq36}
\end{equation}
The velocity component normal to the earth's surface is supposed to be zero.

For a hydrostatic atmosphere and if there is no topography $H = E_i + E_G$, but in that case $E_K$ must be replaced in (\ref{eq36}) by the kinetic energy of horizontal motion $E'_k$:
\begin{equation}
\frac{d}{dt}(A_m + E'_K)
 \: = \:  0
 \: . \label{eq37}
\end{equation}
Under these assumptions the global hydrostatic conservation law (\ref{eq37}) is equivalent to the one derived by Lorenz (1955,1967). 
It is a moist generalization of the dry conservation law 
\begin{equation}
\frac{d}{dt}(A_h + E'_K)
 \: = \:  0
 \: \nonumber
\end{equation}
derived in Marquet (1991).

Similarly to MM the temperature $T_r$ can be defined so that $1/T_r$ is the space-time average over the whole moist atmosphere of $1/T$. 
This choice implies that a uniform (in space and time) heating rate does not generate moist available enthalpy in space-time average. 
All terms in Eq.~(\ref{eq27}) which possess $\eta_T = 1 - T_r/T$ as a multiplicative factor are subject to this property, in particular $(\varepsilon + q)$. 
The value of $T_r$, can be estimated as $250$~K.

If $T_r$ is known, the saturating pressure $p^1_r = p^{31}(T_r)$ and the two energy components $a_T$ and $a_l$ are determined completely at each point of the atmosphere. Following MM the pressure $p^{0}_r = p_r - p^{31}(T_r)$ can then be defined so that the third energy component, $a_p$, cancels out by space-time averaging. 
In this problem the known variables are $m^0, m^t, T_r, p^0$ and $p^1$, and $p^0_r$ is the unknown quantity. 
The value of $p^0_r$ can be estimated as $p_{00}/e \approx 368$~hPa.

Using Eq.~(\ref{eq23}), and the notations of the appendix, these definitions can be summarized
by:
\begin{equation}
\frac{1}{T_r}
\; = \;
\frac{1}{\Delta t} \:
\int^{t_2}_{t_1} \: dt \:
\iiint_{\cal M}
\: \left( \frac{1}{T} \right) \:
\frac{dM}{M}
\label{eq38}
\end{equation}
and
\begin{equation}
\ln\left( p^0_r \right)
\; = \;
\frac{1}{\Delta t} \:
\int^{t_2}_{t_1} \: dt \:
\iiint_{\cal M}
\: \left[ \:
  \frac{m^0}{<m^0>} \: \ln\left( p^0 \right)
   \: + \:
  \frac{m^t}{<m^0>} \: \frac{R^1}{R^0} \:
  \ln\left( \frac{p^1}{p^{31}_r} \right)
\: \right] \:
\frac{dM}{M}
\: ,
\label{eq39}
\end{equation}
where $<m^0>$ is the space-time average of the dry-air concentration (specific content) $m^0$.

 \section{\underline{THE MOIST POTENTIAL TEMPERATURE}.} 
\label{section_8}

The function $a_m$ defined by Eq.~(\ref{eq18}) can be written in an alternative way in order to introduce a moist potential temperature denoted by $\theta^{\ast}$. 
Using a method initiated in HH the sum 
$s - m^0 \: s^0_r - m^t s^1_r$, which is the entropy part of $a_m$, can be expressed in terms of 
\begin{equation}
s \: - \: m^0 \: s^0_r \: - \: m^t s^1_r
\; = \;
m^0 \: c_p^{\ast} \: 
\ln\left( 
 \frac{\theta^{\ast}}{\theta^{\ast}_r} 
\right)
\: , \nonumber
\end{equation}
where $c_p^{\ast} = c_p^0 + r^t \: c_p^1$ and $p_{00}= 10^5$~Pa, leading to
\begin{equation}
a_m
\; = \;
h \: - \: m^0 \: h^0_r \: - \: m^t h^1_r
\: - \: m^0 \: c_p^{\ast} \: T_r \:
\ln\left( 
 \frac{\theta^{\ast}}{\theta^{\ast}_r} 
\right)
\: , \label{eq40}
\end{equation}
where
\begin{equation}
\frac{\theta^{\ast}}{\theta^{\ast}_r} 
\; = \;
\frac{T \: \; (p^0/p_{00})^{-R^0/c_p^{\ast}} \;
           \: (p^1/p_{00})^{-R^1 r^t/c_p^{\ast}} 
     }
     {T_r \: \; (p^0_r/p_{00})^{-R^0/c_p^{\ast}} \;
             \: (p^1_r/p_{00})^{-R^1 r^t/c_p^{\ast}} 
     }
 \: \exp\left( - \:
 \frac{r^2 \: l_{21} \: + \: r^3 \: l_{31}}
      {c_p^{\ast} \: T} 
     \right)
\label{eq41}
\: . 
\end{equation}
This result is obtained starting from Eq.~(\ref{eq18}) of section \ref{section_3.1} written in the form
\begin{equation}
a_m = 
(h - m^0 \: h^0_r - m^t \: h^1_r)
- T_r \: (m^k \: s^k - m^0 \: s^0_r - m^t \: s^1_r) 
\: . \nonumber
\end{equation}
The enthalpy parts $(h - m^0 \: h^0_r - m^t \: h^1_r)$ will not be changed. 
The entropy parts $(m^k \: s^k - m^0 \: s^0_r - m^t \: s^1_r)$ can be transformed using Eq.~(\ref{eq04}) applied to the sum $m^k \: s^k$, yielding 
\begin{equation}
m^0 \: (s^0 - s^0_r) \: + \: m^t \: (s^1 - s^1_r) 
\: + \: m^2 \: (s^2 - s^1) \: + \: m^3 \: (s^3 - s^1) 
\: . \nonumber
\end{equation}
The next step is to express the various differences of entropy, with the introduction of the chemical potentials and the latent heats for $(s^2 - s^1)$ and $(s^3 - s^1)$, leading to
\begin{align}
s^0 - s^0_r & \: = \;
 c_p^0 \: \ln\left( \frac{T}{T_r} \right)
  \: - \:
 R^0 \: \ln\left( \frac{p^0}{p^0_r} \right)
\, , \nonumber \\
s^1 - s^1_r & \: = \;
 c_p^1 \: \ln\left( \frac{T}{T_r} \right)
  \: - \:
 R^1 \: \ln\left( \frac{p^1}{p^1_r} \right)
\, , \nonumber \\
s^2 - s^1   & \: = \;
    \frac{\mu^1 - \mu^2}{T}
    \: - \:
    \frac{h^1 - h^2}{T}
    \; = \;
 R^1 \: \ln\left[ \: \frac{p^1}{p^{21}(T)} \: \right]
    \: - \:
    \frac{l_{21}(T)}{T}
\, , \nonumber \\
s^3 - s^1   & \: = \;
    \frac{\mu^1 - \mu^3}{T}
    \: - \:
    \frac{h^1 - h^3}{T}
    \; = \;
 R^1 \: \ln\left[ \: \frac{p^1}{p^{31}(T)} \: \right]
    \: - \:
    \frac{l_{31}(T)}{T}
\, . \nonumber 
\end{align}
The moist available enthalpy becomes
\begin{align}
\frac{a_m}{T_r} & \: = \;
\frac{h \: - \: m^0 \: h^0_r \: - \: m^t h^1_r}{T_r}
\: - \: 
m^0 \: c_p^{\ast} \: \ln\left( \frac{T}{T_r} \right)
\: + \: 
m^0 \: R^0 \: \ln\left( \frac{p^0}{p^0_r} \right)
\: + \: 
m^0 \: r^t \: R^1 \: \ln\left( \frac{p^1}{p^1_r} \right)
\nonumber \\
 & \; \; \; \; \; \; \; \; \; \; \; \; \; 
   \; \; \; \; \; \; \; \; \; \; \; \; \; \; \; 
\: + \: 
m^0 \: \left(
   \frac{r^2 \: l_{21} \: + \: r^3 \: l_{31}}{T} 
       \right)
\: - \: 
 R^1 \: \left[ \:
   m^2 \ln \left( \frac{p^1}{p^{21}} \right) 
   \: + \:
   m^3 \ln \left( \frac{p^1}{p^{31}} \right) 
       \: \right]
\: . \nonumber
\end{align}
The terms between the square brackets are zero according to Eq.~(\ref{eq08}) and the result is then obtained because Eqs.~(\ref{eq40}) and (\ref{eq41}) are just another way of writing it, by using logarithms and exponentials.

The quantity $\theta^{\ast}$ is analogous from (\ref{eq41}) to the entropic temperature $\theta_S$ of HH, while $\theta^{\ast}$ seems to generalize the ice liquid potential temperature $\theta_{il}$ of Tripoli and Cotton (1981), provided that $\theta^{\ast}_r$ is associated with the denominator of (\ref{eq41}) and if $\theta^{\ast}$ is the product of the numerator and the exponential term.

Using the present notations, $\theta_S$ is defined in HH by
\begin{equation}
s \; = \; m^0 \: s^0_r \: + \: m^t s^2_r
\: + \: m^0 \: 
( c_p^0 + r^t \: c_p^2 ) \: 
\ln\left( 
 \frac{\theta_S}{T_r} 
\right)
\: , \nonumber
\end{equation}
In this study, $\theta^{\ast}$ is similarly related to the entropy by 
\begin{equation}
s \; = \; m^0 \: s^0_r \: + \: m^t s^1_r
\: + \: m^0 \: 
( c_p^0 + r^t \: c_p^1 ) \: 
\ln\left( 
 \frac{\theta^{\ast}}{\theta^{\ast}_r} 
\right)
\: , \nonumber
\end{equation}
As noted in HH, conceptual problems could arise with the $\theta_S$ formulation if liquid water is not present in the moist parcel (i.e. if $r^2 = 0$ but 
$r^t = r^1 + r^3 \neq 0$). 
Indeed it could be difficult to interpret the terms $m^t \: s^2$ and $r^t \: c_p^2$ in this case, whereas in the various terms of (\ref{eq40}) and (\ref{eq41}) only the water vapour element which is always present is involved through $m^t \: h^1_r$, $m^t \: s^1_r$ and $r^t \: c^1_p$. 
This is more consistent.

The advantage of introducing $\theta^{\ast}$ by (\ref{eq40}) lies in a property already mentioned in section \ref{section_3.2}. 
For an adiabatic ($q = 0$) and reversible 
$( \: ( {\mu}^{\alpha} - {\mu}^{\beta} ) \: 
  d_i (m^{\alpha}_{\beta})/dt = 0 \: )$
motion of an inviscid $(\varepsilon = 0)$ and closed 
$(\: d_e(m^k)/dt = 0$ and ${\bf J}^k = 0 \:)$ parcel: 
$d(a_m - h)/dt = - \: T_r \: d(s)/dt = 0$.
This property can be derived from Eqs.~(\ref{eq24}) and (\ref{eq30}) together with the remarks made at the beginning of section \ref{section_6.1}:
\begin{eqnarray} 
\!\!\!\!\!\!\!\!\!\!\!
0 \; = \;
 \frac{T}{T_r} \; \frac{d\, (a_m - h)}{dt}  & = & 
           \: - \: ( \varepsilon  + q )
           \: + \: \frac{1}{\rho} \; {\bf J}^k  . \: {\bf \nabla} ( h^k )
   \: - \: T \: s^k \: \frac{d_e \, m^k }{dt}
   \: + \: \frac{T}{T_r} \: (\mu^0_r - \mu^1_r) \: \frac{d_e \, m^t}{dt}
\nonumber \\ 
\!\!\!\!\!\!\!\!\!\!\!
& &
    + \: \left\{ \:  
           (\mu^1 \: - \: \mu^2 ) \: \frac{d_i \, m^1_2}{dt}
           \: + \:
           (\mu^2 \: - \: \mu^3 ) \: \frac{d_i \, m^2_3}{dt}
           \: + \:
           (\mu^1 \: - \: \mu^3 ) \: \frac{d_i \, m^1_3}{dt}
    \: \right\}
 . \label{eq42}
\end{eqnarray}
For such a motion, the quantities ``$a_m - h$'', ``$m^0 \: h^0_r + m^t \: h^1_r$'', ``$m^0 \: c_p^{\ast} \: T_r$'', and ``$\theta^{\ast}_r$'' are all constants and one can deduce from Eqs.~(\ref{eq40}) to (\ref{eq42}) the conservation of
\begin{equation}
\theta^{\ast}
\; = \;
T \: \left(\frac{p^0}{p_{00}}\right)^{-R^0/c_p^{\ast}}
  \: \left(\frac{p^1}{p_{00}}\right)^{-R^1 r^t/c_p^{\ast}} 
  \: \exp\left( - \:
 \frac{r^2 \: l_{21} \: + \: r^3 \: l_{31}}
      {c_p^{\ast} \: T} 
     \right)
\label{eq43}
\end{equation}
during the motion of the parcel. This conservative property holds whatever the state of the parcel or its evolution with time (saturated or not, with liquid water or ice). The quantity $\theta^{\ast}$ can be called the moist potential temperature associated with the moist entropy, $s$, and with the moist available enthalpy, $a_m$

 \section{\underline{CONCLUDING REMARKS}.} 
\label{section_9}

In this paper the moist available enthalpy $a_m$ is defined as a local exergy-like function.
The concept of potential change in total entropy has been used to generalize the dry
available enthalpy $a_h$ to the moist atmosphere case, different from the real exergy
approach where a Kullback function can be used. A moist reference dead state is
introduced to define $a_m$ but the name ``reference'' does not here have the same meaning
as it is understood in the meteorological approach of available energy. It is not a reference
state which could be constructed from a given state of the atmosphere neither to minimize
the total moist enthalpy (Lorenz), nor to maximize the total entropy with some constraints
(Dutton). It is a chemically stable local state, a passive environment, and any parcel of
the real atmosphere can reach this reference dead state separately from each other.

The mathematical expression of $a_m$ is simple and analytical, this corresponds to
Lorenz's (1978, 1979) demand. In fact it is possible to put $a_m$ into various synonymous
forms, such as Eqs. (17), (18), (19) and (40). Each of these forms yields a special property
satisfied by $a_m$. 
\begin{itemize}[label=$\bullet$,leftmargin=3mm,parsep=0cm,itemsep=0.1cm,topsep=0cm,rightmargin=2mm]
\vspace*{-1mm}
\item  
Equation (17) means that $a_m$ is independent of any arbitrary choice for
the absolute values of energy, enthalpy or entropy functions. 
\item  
Equation (18) shows that $a_m$ is similar to what is called flow-exergy in thermodynamics. 
\item  
Equation (19) is the separation of $a_m$ into the three components $a_T$, $a_l$ and $a_p$. 
These components are the generalizations of the same components already defined with 
the dry available enthalpy, except for the latent part, $a_l$, which is a new one. 
\item  
Equation (40) serves as an introduction for the moist potential temperature, ${\theta}^{\ast}$, 
which satisfies conservative properties.
\end{itemize}

Moreover, using the local-state theory of moist thermodynamics, the local function
$a_m$ leads to:
\begin{itemize}[label=,leftmargin=3mm,parsep=0cm,itemsep=0.1cm,topsep=0cm,rightmargin=2mm]
\vspace*{-1mm}
\item  
(i) a moist Bernoulli law satisfied by $a_m + e_G + e_K$;
\item  
(ii) a moist local energy cycle involving $a_m$, $e_G$ and $e_K$; and
\item  
(iii) a moist integral hydrostatic conservative property satisfied by $A_m + E'_K$.
\end{itemize}

Another property is that $a_m$, like $a_h$ in the dry study (Marquet, 1991), is only partially
convex. Indeed, only ${\cal F}(X)$ and $a_T$ are positive and convex in $X$. 
On the contrary, for an
ideal gas, the exergy function 
$[\: {e_i - (e_i)_r} \:] + p_r  \:[ \: {(\rho)^{-1} - ({\rho}_r)^{-1}} \: ]- T_r \: (s - s_r)$
is doubly convex, since it can be rewritten as the sum of two terms depending 
on the function ${\cal F}(X) =  X - \ln(1 + X)$: 
$c_p \: T_r \:{\cal F} (T/T_r - 1) + R \: T_r \: {\cal F}({\rho}_r/{\rho} - 1)$. 
The possibility of cancelling out the
integral of $a_p$ is used in this study to define the value of $p_r$, 
it is not possible to have this result with a convex function.

The purpose of the present paper is to show that the exergy-like function $a_m$ verifies
{\it at the same time \/} all the properties recalled in this section. It is a general thermodynamic
quantity which was not easy to guess from the two laws of thermodynamics, although
the moist version (Eqs. (19) to (23)) requires only modest modification of the expression
for the dry case.

The author is aware of the theoretical nature of the present paper. At this point
numerical evaluations of the various concepts and functions introduced in this study must
still be realized with case studies, for instance for an open moist synoptic-scale system.
It is a knowledge of the conversion terms which would be the more revealing of the
interest of this exergetic analysis. An attempt is, however, made in section 4 to estimate
the orders of magnitude of the moist available enthalpy components with a vertical profile
that simulates a cumulus cloud. It is found that the moist contributions are important in
comparison with the dry version. The results of a synoptic-scale study using dry and
moist versions of the available enthalpy concept will be presented in a future paper. The
dry and moist exergy cycles will be applied with the hydrostatic assumption, stressing
the problem of the various energy conversions between isobaric layers of an open limited
atmospheric domain, including the study of boundary fluxes.

The various results derived in this paper can, however, be applied from now on.
The first step is to determine the two constants $T_r$ and $p_r$ according to 
the definitions (38) and (39). 
It is also possible to use prescribed realistic values such as $251$~K and $370$~hPa.
The moist available enthalpy components can then be computed from Eqs. (19) to (23).
The budget equation for $a_m$ can be studied starting from Eq. (30), or with the suitable
assumptions from the simplified Eq. (34). The simplified moist available enthalpy cycle
is not very different from other enthalpy cycles and it is not very interesting in this form,
except it was necessary to demonstrate that this cycle exists with the moist formulation.
On the contrary the Bernoulli equation (35) can be useful in order to develop some
qualitative or quantitative arguments such as Convective Available Potential Energy
considerations or Carnot heat-engine analogy applied to convection or cyclone problems
for instance. Finally, the conservative properties of the moist potential temperature
defined by Eqs. (40), (41) and (43), which solely come from the second law of thermodynamics,
can serve as accurate air-mass analysis.

On a more general point of view, it could be interesting to find a function of statistical
physics similar to the Kullback function in order to provide another consistent physical
meaning to $a_m$. It would also be worth taking into account precipitation which falls at
$T'_w \neq T $ in unsaturated air. This implies generalizing $a_m$  using, for instance, thermodynamics
of a thermically heterogeneous medium. Finally, we may ask whether these
moist functions could be used as moist prognostic variables in numerical models dealing
with liquid water or ice.

\vspace{5mm}
\noindent{\large\bf Acknowledgements}
\vspace{2mm}

This work has been partially performed during a two-year post-engineering-school
stay at the Laboratoire de M\'et\'eorologie Dynamique, Ecole Polytechnique, France. 
I thank Dr D. L. Cadet for his support during this study and I am also grateful to J. F.
Geleyn and J. F. Royer for their helpful suggestions. Thanks are also due to R. P. Pearce
and the anonymous reviewer for their constructive comments.

\newpage
\noindent
{\large\bf Appendix A. List of symbols.}
             \label{appendixSymbol}
\renewcommand{\theequation}{A.\arabic{equation}}
  \renewcommand{\thefigure}{A.\arabic{figure}}
   \renewcommand{\thetable}{A.\arabic{table}}
      \setcounter{equation}{0}
        \setcounter{figure}{0}
         \setcounter{table}{0}
\vspace{1mm}
\hrule

\vspace{4mm}
{\bf \large General remarks}
\vspace{1mm}

The subscript $r$ denotes the values in the reference dead state (like in $p_r$ and $T_r$).
The subscripts or superscripts $i, j . k, \alpha$ and $\beta$ are used to denote partial values: $0$ for dry air, $1$ for water vapour, $2$ for liquid water (stable or supercooled) and $3$ for ice 
(see Hauf and H\"oller, 1987). 
When they are repeated twice in two different terms, these subscripts or superscripts represent an implicit sum over the values $0$ to $3$ 
(Einstein's notation, with for example: $a_m = m^k \: a_m^k = 
\sum^3_{k=0} m^k \: a_m^k $).

\vspace{4mm}
{\bf \large Basic thermodynamics}
\vspace{0mm}

\begin{tabbing}
 --------------------------------\=  ------------------------------------------------------------------------- --\= \kill
 $T$, $p$, $\theta$ \> Temperature, total pressure, potential temperature \\
 $h$, $e_i$, $s$ \> Specific enthalpy, internal energy, entropy \\
 $H$, $E_i$, $S$ \> Integrated forms of $h$, $e_i$, $s$ \\
 $\rho^k$, $\rho = \sum^3_{k=0} \rho^k$ \>  Partial and total densities \\
 $v = 1/\rho$ \> Specific volume \\
 $m^k = \rho^k/\rho$ \> Concentrations or specific contents $q_k$ ($k=0$ to $3$) \\
 $m^t = m^1 + m^2 + m^3$ \> Total-water concentration (or specific content, 
                            per unit mass of moist air) \\
 $r^k = m^k / m^0$ \> Mixing ratios of water species ($k=1$ to $3$) \\
 $r^t = r^1 + r^2 + r^3$ \> Total-water mixing ratio of the water element \\
 $p^0$, $p^1$ \> Partial pressures for dry air and water vapour 
                 ($p = p^0 + p^1$) \\
 $R^i$ ($i=0$ and $1$) \> Gas constants: $R = R^0 + R^1$\\
 $c_p^j, c_v^j$ ($j=0$ to $3$) \> Specific heats at constant pressure and volume, \\
  \> with $c_p^j = c_v^j = c^j$ for the condensed phases ($j=2$ and $j=3$) \\
 $c_p = m^k \: c_p^k$ \>  Total specific heats at constant pressure
                          ($= \sum^3_{k=0} c_p^k$) \\
 $c_p^{\star} = c_p^0 + r^t \, c_p^1$ \>  A notation used in sections 7 and 3.3 \\
 $T_{tr} = 273.16$~K \> Triple-point temperature \\
 $l_{ij} = h^j - h^i$ \> Latent heat ($i$ to $j$ transformation)\\
 $T'_{w}$ \> Wet-bulb temperature \\
 $p^{i1}(T)$ \> Saturating pressure over liquid water ($i=2$) and ice ($i=3$) \\
 $\mu^k = h^h - T \: s^k$ \> Chemical potentials (or Gibbs function) of
            the element $k=0$ to $3$ \\
 ${\cal A}^i = \mu^1 - \mu^i$ \> Chemical affinities related to
                $i$ form ($i=2$ and $3$) and water vapour ($i=1$) \\
\end{tabbing}

\vspace{2mm}
{\bf \large Basic kinematics}
\vspace{0mm}

\begin{tabbing}
 --------------------------------\=  ------------------------------------------------------------------------- --\= \kill
 $e_K$, $e_G$ \> Specific kinetic energy (of 3D wind) and potential energy \\
 $E_K$, $E_G$ \> Integrated forms of $e_K$ and $e_G$ \\
 $E'_K$        \> Integrated kinetic energy of the horizontal wind \\
 $z$, $\omega$ \> Upward distance, vertical wind component in 
                 isobaric coordinate \\
 $\overline{\omega}$; $\overline{T}$ \> 
        Isobaric average of $\omega$ and $T$ \\
 $\partial/\partial t$, $d/dt$ \> Eulerian derivative, Lagrangian derivative 
                (or for a function of $t$ only) \\
 $d_e/dt$, $d_i/dt$ \> External and internal Lagrangian change in time \\
 $d_i(m^{\beta}_{\alpha})/dt$ \> Chemical Lagrangian rate of change (of phases)
                    of $\alpha$ into $\beta$ element \\
 ${\bf v}^k$ \> Partial velocity of the element $k$ in the fluid parcel \\
 ${\bf v} = \rho^k \: {\bf v}^k / \rho$ \>  
              Barycentric velocity of the parcel \\
 ${\bf J}^k$ \> Diffusion flux of the element $k$\>  \\
 ${\cal J}^k $ \> Phase flux of the element $k$ \\
 ${\bf J}_{iq}$ \>  The inner heat flux\\
 $\boldmath{\nabla}$ \> The gradient (or Del, or Nabla) operator \\
 $q = - (\rho)^{-1} \: \boldmath{\nabla} \, . \, {\bf J}_{iq} $ \> 
             The specific (per unit mass of moist air) heating rate \\
 $\boldmath{\sigma}$ \> Viscous stress tensor \\
 $\varepsilon = (\rho)^{-1} \: (\boldmath{\sigma} 
      \: . \boldmath{\nabla}) \: . \: {\bf v} $ \> Positive Rayleigh function \\
 ${\bf g}$ \> The gravity force (vector) \\
 ${\bf k}$ \> The unit upward vertical vector \\
 $g = - \: {\bf g} \: . \: {\bf k}$ \> The magnitude of gravity (scalar) \\
 $\phi$ \> The geopotential ($=g \: z$) \\
 ${\bf F}_r$ \> External forces \\
 $C_{(G,K)}$, $C_{(m,K)}$ \> Conversion terms between $e_G$ and $e_K$,
       and between $a_m$ and $e_K$, respectively \\
\end{tabbing}

\vspace{4mm}
{\bf \large Exergy theory}
\vspace{0mm}

\begin{tabbing}
 --------------------------------\=  ------------------------------------------------------------------------- --\= \kill
 $p_r = p^0_r + p^1_r$, $T_r$ \> Reference ``dead state'' pressures 
        and temperature (moist case) \\
 $P_r$ \> Reference pressure for the dry-air enthalpy case \\
 $a_h$, $a_m$ \> Specific dry and moist available enthalpy \\
 $a_m^k = T_r \: (\Delta_mS^o)^k$ \> 
          The partial moist available enthalpy ($k=0$ to $3$), 
          $a_m=m^k \: a_m^k$  / the ``change \\
      \>  in total entropy'' or ``surface of dissipated energy''
          vision of Gibbs (1873) \\
 $b^k$ \> A notation used in section 3.1 for 
            $(h^k - h^k_r) - T_r \: (s^k - s^k_r)$ \\
 $a_e$ \> The specific available energy \\
 $a_T, a_l, a_p$ \> Specific temperature, latent (water) 
                   and pressure components of $a_m$ \\
 $A_h, A_m, A_T, A_l, A_p$ \> Integrated forms of 
                   $a_h, a_m, a_T, a_l, a_p$ \\
 $T_0 \: \Sigma$ \> The static entropic energy (notations of 
                    Dutton, Johnson, Livesey, Pichler) \\
 $\theta^{\ast}; \theta^{\ast}_r$ \> 
         The moist-entropy potential temperatures 
         (real and reference values) \\
 $\theta_S$ \> 
         The moist-entropy potential temperatures 
         in Hauf and H\"oller (1987) \\
 ${\cal F}(X) = X - \ln(1+X)$ \> 
         : the positive function used in the temperature moist-air
           component $a_T$ \\
         \> (see the section 3.3, where $X=(T-T_r)/T_r=T/T_r -1$) \\
 $\eta_T = (T-T_r)/T_r$ \> The (Carnot) efficiency factor
               (also equal to $1- T_r/T$) \\
 $R_{max}$ \> The ``maximum delivered work'', or ``available energy'' 
              (Landau and Lifchitz, \\
           \> 1976, $\S$20, except they use $R_{min}$ to denote the 
           maximum delivered work as \\
           \> a ``minimum received work) \\
 $T_0$ \>  Temperature of the thermostat \\
 $T_1$ \>  Equilibrium temperature used in the definition of $a_e$, 
           see section 3.1\\
 $\Delta_1 S_t$ and $\Delta_2 S_t$ \> 
          Two changes in total entropy used in the definition of $a_e$, 
           see section 3.1 \\
 $\Delta S_t = R_{max} / T_0$ \> 
          The difference in total entropy between the real state and
          the associated \\
          \> equilibrium state of temperature $T_0$ 
          (see Landau and Lifchitz, 1976, $\S$20)\\
 $(E_t)$ \> The ``total'' energy (for a ``unit mass parcel'' 
          plus a ``thermostat'') \\
 $(S_t)$ \> The ``total'' entropy (for a ``unit mass parcel'' 
          plus a ``thermostat'') \\
 $\Delta S$ \> The change in ``total'' entropy for any process
          involving the system \\
          \> ``unit mass parcel''  plus ``thermostat'' 
          (Gibbs, Landau and Lifchitz)\\
 $(S_t)_{eq}$ \> The ``equilibrium'' curve (Gibbs, Landau and Lifchitz) \\
 $\Delta_d S^o$ \> The dry-air potential change in total entropy \\
 $\Delta_m S^o$ \>  The moist (total, or partial for $k=0$ to $3$)
                    potential change in total entropy \\
\end{tabbing}

\vspace{4mm}
{\bf \large Other quantities}
\vspace{0mm}

\begin{tabbing}
 --------------------------------\=  ------------------------------------------------------------------------- --\= \kill
 $p_{00} = 10^5$~Pa \> A conventional pressure \\
 ${\cal H}_n$ \> A label for the hypothesis number $n$ \\
 $H^k$ \> An equivalent height \\
 $t_2, t_1, \Delta t = t_2-t_1$ \> Times, a time interval \\
 $<m^0>$ \> The space-time average of the dry-air concentration $m^0$
          (specific content) \\
 $dM, M$ \> An element of mass, the mass of the atmosphere \\
 ${\cal M}$ \> The mass integrating domain of the atmosphere \\
 $\psi$ \> A dummy specific function \\
\end{tabbing}


\vspace{5mm}
\noindent{\Large\bf References}
\vspace{2mm}

\noindent{$\bullet$ Ahrendts, J.} {(1980)}
Reference states. 
{\it Energy\/}, {\bf 5}, 667--677

\noindent{$\bullet$ Blackburn, M.} {(1983)}
{\it An energetic analysis of the general atmospheric circulation\/}.
Department of Meteorology thesis. 
University of Reading.

\noindent{$\bullet$ Brennan, F. E., Vincent, D. G.} {(1980)}
Zonal and eddy components of the synoptic-scale energy
budget during intensification of hurricane Carmen (1974).
{\it Mon. Weather Rev.\/}, {\bf 108}, 954--965

\noindent{$\bullet$ De Groot, S. R. and Mazur, P.} {(1962)}
{\it Non-equilibrium thermodynamics\/}. 
North-Holland Pub. Company, Amsterdam

\noindent{$\bullet$ Dutton, J . A.} {(1973)}
The global thermodynamics of atmospheric motion. 
{\it Tellus\/}, {\bf 25}, (2). 89-110

\noindent{$\bullet$ Dutton, J . A.} {(1976)}
{\it The ceaseless wind\/}. McGraw-Hill

\noindent{$\bullet$ Dutton, J. A. and Johnson, D. R.} {(1967)}
{\it The theory of available potential energy and a variational
approach to atmospheric energetics\/}. 
Pp. 333-436, in Vol.12 of Advance in geophysics. 
Academic Press, New York and London

\noindent{$\bullet$ Evans, R. B.} {(1969)}
{\it A proof that exergy is the only consistent measure 
of potential work\/}. 
Doctor of Philosophy thesis. 
Thayer School of Engineering, Dartmouth College, 
Hanover

\noindent{$\bullet$ Evans, R. B.} {(1980)}
Thermoeconomic isolation and essergy analysis. 
{\it Energy\/}, {\bf 5}, 805--821

\noindent{$\bullet$ Gibbs. J. W.} {(1873)}
A method of geometrical representation of the thermodynamic
properties of substance by means of surfaces. 
{\it Trans.
Connecticut Acad.\/}, {\bf 2}, 382404. (Pp. 33-54 in Vol. 1 of
{\it The collected works of J. W. Gibbs\/}, 
1928. Longmans Green and Co.)

\noindent{$\bullet$ Glansdorff, P. and Prigogine. I.} {(1971)}
{\it Structure stabilit\'e et fluctuations.\/} 
Masson et Cie, Paris

\noindent{$\bullet$ Hauf, T. and H\"oller, H.} {(1987)}
Entropy and potential temperature. 
{\it J. Atmos. Sci.\/} {\bf 44}, (20). 2887--2901

\noindent{$\bullet$ Haywood, R. W.} {(1974)}
A critical review of the theorems of thermodynamic availability,
with concise formulations. 
{\it J. Mech. Eng. Sci.\/} 
Part~1: Availability, 16, (3) 160-173, 
Part~2: Irreversibility, {\bf 16}, (4), 256--267.

\noindent{$\bullet$ Johnson, D. R. and Downey, W. K.} {(1982)}
On the energetics of open systems. 
{\it Tellus\/}, {\bf 34}, (5), 458--470

\noindent{$\bullet$ Karlsson. S.} {(1990)}
{\it Energy, entropy and exergy in the atmosphere\/}. 
Institute of Physical Resource Theory thesis. 
Chalmers University of Technology, 
G\"oteborg, Sweden

\noindent{$\bullet$ Kestin. J.} {(1980)}
Availability: the concept and associated terminology. 
{\it Energy\/}. {\bf 5}, 679--692

\noindent{$\bullet$ Kullback. S.} {(1959)}
{\it Information theory and statistics\/}, 
John Wiley \& Sons, New York and London

\noindent{$\bullet$ Landau. L. D. and Lifchitz, E. M.} {(1976)}
{\it Statistical physics} (third edition). 
Pergamon Press, London

\noindent{$\bullet$ Lebon. G. and Mathieu, Ph.} {(1981)}
\'Etude compar\'e de diverses th\'eories de thermodynamique du
non-\'equilibre. 
{\it Entropie\/}, {\bf 100}, 75-86

\noindent{$\bullet$ Livezey, R. E. and Dutton, J . A.} {(1976)}
The entropic energy of geophysical fluid systems. 
{\it Tellus\/}. {\bf 28}, (2), 138--157

\noindent{$\bullet$ Lorenz. E. N.} {(1955)}
Available potential energy and the maintenance of the general
circulation.
{\it Tellus\/}, {\bf 7}, (2), 157-167

\noindent{$\bullet$ Lorenz. E. N.} {(1967)}
{\it The nature and theory of the general circulation of the 
atmosphere\/}. 
World Meteorological Organization

\noindent{$\bullet$ Lorenz. E. N.} {(1978)}
Available energy and the maintenance of a moist circulation.
{\it Tellus\/}, {\bf 30}, (1), 15-31

\noindent{$\bullet$ Lorenz. E. N.} {(1979)}
Numerical evaluation of moist available energy. 
{\it Tellus\/}, {\bf 31}, (3), 230--235

\noindent{$\bullet$ McHall. Y. L.} {(1990a)}
Available potential energy in the atmosphere. 
{\it Meteorol. Atmos. Phys.\/}, {\bf 42}, 39-55

\noindent{$\bullet$ McHall. Y. L.} {(1990b)}
Generalized available potential energy. 
{\it Adv. Almos. Sci.\/}, {\bf 7}, (4). 395-408

\noindent{$\bullet$ McHall. Y. L.} {(1991)}
Available equivalent potential energy in moist atmospheres.
{\it Meteorol. Atmos. Phys.\/}, {\bf 45}, 113-123

\noindent{$\bullet$ Margules, M.} {(1905)}
{\it On the energy of storms\/}, Smithsonian miscellaneous collections,
{\bf 51}, 4, 533-595, 1910 (Translation by C. Abbe
from ``\"Uber die energie der st\"urme''. 
{\it Jahrb. Zentralantst. Meteorol.\/} {\bf 40}, 1--26, 1905)

\noindent{$\bullet$ Marquet, P.} {(1990a)}
La notion d'enthalpie utilisable: application \`a 
l'\'energ\'etique atmosph\'erique. 
{\it C. R. Acad. Sci. Paris\/}, {\bf 310}, s\'erie~II, 1387--1392.

\noindent{$\bullet$ Marquet, P.} {(1990b)}
L'exergie de l'atmosph\`ere. D\'efinition et propri\'et\'es 
de l'enthalpie utilisable humide. 
Pp. 81-86 in
{\it Atelier de mod\'elisation de l'atmosph\`ere\/}. 
Available from: M\'et\'eo-France.
C.N.R.M. 31057 Toulouse Cedex, France

\noindent{$\bullet$ Marquet~P.} {(1991)}.
{On the concept of exergy and available
enthalpy: application to atmospheric energetics.
{\it Q. J. R. Meteorol. Soc.\/}
{\bf 117}:
449--475.
\url{http://arxiv.org/abs/1402.4610}.
{\tt arXiv:1402.4610 [ao-ph]}
}

\noindent{$\bullet$ Meixner, J. and Reik, H. G.} {(1959)}
Thermodynamik der irreversiblen prozesse. 
{\it Enc. of Physics\/}, III/2. 
Springer, Berlin

\noindent{$\bullet$ Michaelides, S. C.} {(1987)}
Limited area energetics of Genoa cyclogenesis. 
{\it Mon. Weather Rev.\/}, 
{\bf 115}, 13--26

\noindent{$\bullet$ Morris, D. R. and Szargut, J.} {(1986)}
Standard chemical exergy of some elements and compounds
on the planet Earth. 
{\it Energy\/}, {\bf 11}, (8). 733--755

\noindent{$\bullet$ Muench, H. S.} {(1965)}
On the dynamics of the wintertime stratosphere circulation. 
{\it J. Atmos. Sci.\/}, {\bf 22}, 349--360

\noindent{$\bullet$ Pearce, R. P.} {(1978)}
On the concept of available potential energy. 
{\it Q. J. R. Meteorol. Soc.\/}, 
{\bf 104}, 737--755

\noindent{$\bullet$ Pichler, H.} {(1977)}
Die bilanzgleichung f\"ur die statischer entropische Energie der
Atmosph\"are. 
{\it Arch. Met. Geophys Biokl.\/}, 
Ser.~A, {\bf 26}, 341--347

\noindent{$\bullet$ Pontaud, M., Marquet, P., Bernard-Boussiere, I., Muzellec, A, Vincent, C. and Finet, A.} {(1990)}
R\'esultats num\'eriques d'un cycle \'energ\'etique 
local appliqu\'e \`a une perturbation’. 
Pp.~87-92 in 
{\it Atelier de modelisation de l'atmosph\`ere\/}. 
Available from: M\'et\'eo-France, C.N.R.M.
31057 Toulouse Cedex, France

\noindent{$\bullet$ Prigogine, I.} {(1949)}
Le domaine de validit\'e de la thermodynamique des ph\'enom\`enes
irr\'eversibles. 
{\it Physica\/}, {\bf 15}, 272--284

\noindent{$\bullet$ Szargut , J.} {(1980)}
International progress in second law analysis. 
{\it Energy\/}. {\bf 5}, 709--718

\noindent{$\bullet$ Szargut, J. and Styrylska, T.} {(1969)}
Die exergetische analyse von prozessen der feuchten luft. 
{\it Heiz. L\"uft. Haustechn.\/}, {\bf 20}, (5). 173--178

\noindent{$\bullet$ Thomson, W.} {(1853)}
On the restoration of mechanical energy from an unequally
heated space. 
{\it Phil. Mag.\/}, {\bf 5}, 4e series, 102--105

\noindent{$\bullet$ Thomson, W.} {(1879)}
On thermodynamic motivity. 
{\it Phil. Mag.\/}, {\bf 7}, 5e series, 348--352

\noindent{$\bullet$ Tripoli, G. J. and Cotton, W. R.} {(1981)}
The use of ice-liquid water potential temperature 
as a thermodynamic variable in deep atmospheric models. 
{\it Mon. Weather Rev.\/}, {\bf 109}, 1094--1102

\noindent{$\bullet$ Van Mieghem, J.} {(1956)}
The energy available in the atmosphere for conversion 
into kinetic energy. 
{\it Beitr. Phys. Atmos.\/}, {\bf 29}, 129--142

  \end{document}